\documentclass[onecolumn]{elsart3p}
\journal{Journal of Computational Physics}

\usepackage{algorithm,algorithmic}
\usepackage{amsmath,amssymb,amsfonts}
\usepackage{graphics,graphicx}
\usepackage{multirow}
\usepackage{color}
\usepackage{tikz}
\usepackage{setspace}

\newcommand{\uA}{\mathcal{A}}
\newcommand{\bA}{{\bf A}}
\newcommand{\bB}{{\bf B}}
\newcommand{\bC}{{\bf C}}
\newcommand{\bL}{{\bf L}}
\newcommand{\bM}{{\bf M}}
\newcommand{\bO}{{\bf O}}
\newcommand{\bKron}{{\bf \delta}}
\newcommand{\bG}{{\bf G}}
\newcommand{\bQ}{{\bf Q}}
\newcommand{\bPsi}{{\bf \Psi}}

\newcommand{\iA}{{\it A}}

\newcommand{\Ewr}{\mathcal{E}_{r}}
\newcommand{\Ewk}{\mathcal{E}_{k}}
\newcommand{\rv}{{\bf b}}
\newcommand{\lv}{{\bf a}}
\newcommand{\bk}{\vec {\bf k}}
\newcommand{\bn}{\vec {\bf n}}
\newcommand{\br}{\vec {\bf r}}
\newcommand{\bR}{\vec {\bf R}}
\newcommand{\bt}{\vec {\bf t}}
\newcommand{\Lmu}{\mathcal{L}_\mu}
\newcommand{\Lma}{\mathcal{L}_1}
\newcommand{\Lmb}{\mathcal{L}_2}
\newcommand{\Lmc}{\mathcal{L}_3}

\newcommand{\RRR}{\mathbb{R}^3}
\newcommand{\order}{\mathcal{O}}

\begin{document}
\begin{frontmatter}
\title{An $\mathcal{O}(N)$ Method for Rapidly Computing Periodic Potentials Using Accelerated Cartesian Expansions}

\author[ECE,PA]{A.D. Baczewski},
\ead{baczewsk@msu.edu}
\author[ECE,PA,cor]{B. Shanker},
\ead{bshanker@msu.edu}

\address[ECE]{Michigan State University, Department of Electrical and Computer Engineering \\ East Lansing, Michigan, USA}
\address[PA]{Michigan State University, Department of Physics and Astronomy \\ East Lansing, Michigan, USA}
\corauth[cor]{Corresponding author.}

\begin{abstract}

The evaluation of long-range potentials in periodic, many-body systems arises as a necessary step in the numerical modeling of a multitude of interesting
physical problems. Direct evaluation of these potentials requires $\mathcal{O}(N^2)$ operations and $\mathcal{O}(N^2)$ storage, where $N$ is the number 
of interacting bodies.  In this work, we present a method, which requires $\mathcal{O}(N)$ operations and $\mathcal{O}(N)$ storage, for the evaluation of 
periodic  Helmholtz, Coulomb, and Yukawa potentials with periodicity in 1-, 2-, and 3-dimensions, using the method of Accelerated Cartesian Expansions (ACE).  
We present all aspects necessary to effect this acceleration within the framework of ACE including the necessary translation 
operators, and appropriately modifying the hierarchical computational algorithm. We also present several results that validate the efficacy of this method 
with respect to both error convergence and cost scaling, and derive error bounds for one exemplary potential. 

\end{abstract}

\end{frontmatter}
\maketitle

\doublespacing

\section{Introduction}

The evaluation of potential functions in many-body systems subject to periodic or quasi-periodic boundary conditions is a computationally demanding task that arises frequently in the numerical modeling of physical systems.  Among 
the many contexts in which such calculations arise are the analysis of electromagnetic wave propagation in photonic bandgap structures \cite{otani08a,otani09}, frequency selective structures \cite{munk00}, cosmological structure 
formation \cite{springel05a}, defects in the solid state \cite{hung09,shin09}, etc.  Numerical methods specific to the solution of these types of problems require the repeated evaluation of periodic potentials, whether in the 
application of an iterative solver, or in the step-by-step updating of energies and force fields in a time integration scheme. It is well-known that direct approaches to the evaluation of periodic potentials require $\order(N^2)$ 
operations and $\order(N^2)$ storage, where $N$ is the number of unknown quantities in a single unit cell of the periodic structure. Consequently, there is a need for the development of fast methods that mitigate this 
quadratic scaling. 

While this work is concerned with such methods for periodic/quasi-periodic potentials, these algorithms have been researched extensively for non-periodic potentials, and we first discuss 
the more general aspects of fast methods for arbitrary pairwise potentials.  The history of such methods spans more than four decades, and during this period a number of distinct algorithms have been developed. 
They can be broadly categorized by whether acceleration is achieved through a hierarchical decomposition of the domain, as in tree codes \cite{barnes86,duan00} and fast multipole methods (FMMs) \cite{greengard87,schmidt91}, 
or the discretization and resolution of the potential across multiple scales, as in particle-mesh \cite{hockney_book} and multigrid methods \cite{izaguirre05}. Algorithms from the latter category predate hierarchical methods 
by more than a decade, and have matured considerably with time.  The basic premise of particle-mesh/multigrid methods is the evaluation of the potential based upon data defined on a hierarchy of discretization scales.  A global 
solution is generated rapidly at the coarsest scale, and local corrections are then generated and applied based upon data at the finest scale. Perhaps the oldest of these methods is particle-particle/particle-mesh (P$^3$M), 
first published in 1973 in the context of molecular dynamics simulations \cite{hockney73}.  The basis of P$^3$M is the interpolation of point sources onto a mesh at a coarser scale (particle-mesh), from which a continuum source 
distribution can be defined, at which point the potential is computed using an FFT-based PDE solver for the associated continuum problem.  While this reproduces the potential accurately due to long-range interactions, information 
about short-range interactions is lost in interpolation, and corrections are subsequently applied by way of the direct (particle-particle) evaluation of the potential due to point sources which are in close spatial proximity. 
Some other, closely related methods include particle-mesh Ewald (PME), smooth PME, and multigrid methods.  Typically, these methods require $\order(N \log N)$ overhead in terms of both number of operations and storage.  An 
extensive list of references concerning mesh-based methods can be found in \cite{griebel09}. 

In spite of their frequent conflation in the literature, tree codes and FMMs represent two distinct approaches to the efficient evaluation of pairwise potentials that are best explained in \cite{cheng99}.  The first 
tree codes, due to Barnes and Hut \cite{barnes86}, were based on the observation that the gravitational/Coulombic potential evaluated at a point due to sources far away can be accurately represented in terms of a truncated 
multipole expansion of the source distribution.  More generally, in tree codes the interaction between a source-observer pair is computed using one of three options: (i) directly, (ii) at each observation point using 
the multipole expansion due to a cluster of sources, or (iii) using local expansions at a cluster of observers.  The decision as to which operation is to be used is made to guarantee computational efficiency.
Given the generality of presentation, it is worth noting that tree codes have been developed for a variety of potentials \cite{duan00,li09}. FMMs take this principle one step further, by introducing aggregation and disaggregation 
operators \cite{greengard87}, that permit the computation of potentials in a completely optimal manner \cite{cheng99}.  In general, these methods lead to a cost complexity with $\order(N \log^{\alpha} N)$ scaling, where 
$\alpha \in \left[0,1\right]$ depends upon the distribution of unknowns. Among the principle advantages of tree/fast multipole methods is that they have mathematically rigorous bounds on error, which are often lacking in 
mesh-based methods. 

\subsection{Accelerated Cartesian Expansions - A brief introduction}

The method of Accelerated Cartesian Expansions (ACE) is a tree-based method similar in spirit to FMMs, in so far as it includes aggregation and disaggregation operators. Classical FMMs rely on constructing a representation 
of the Green's function in terms of special functions; for instance, the Coulombic Green's function is represented in terms of spherical harmonics.  The ACE algorithm, on the other hand, is contingent upon constructing a 
representation in terms of a generalized Taylor expansion that is expressed in terms of totally symmetric tensors which are used to reduce the overall cost relative to other Cartesian methods.  In 2007, ACE was introduced for the evaluation of
potentials of the form $r^{-\nu}$ \cite{shanker07}. While it is based on Taylor series expansions, it was shown that it is possible to develop {\em exact} aggregation and disaggregation operators. More interestingly, using the well-known equivalence 
between traceless Cartesian tensors and Legendre polynomials, it was shown that it is possible to develop relationships, both in terms of cost and operations, between ACE and classical FMM. As ACE is not wedded to addition theorems 
for special functions, it is possible to apply these to the rapid evaluation of many different potentials. To date, this has been done for potentials of the form $r^{-\nu}$ ($\nu \in \mathbb{R}$) \cite{shanker07}, 
Lienard-Wiechert potentials \cite{vikram07}, and diffusion, Klein-Gordon, and lossy wave potentials \cite{vikram10}.  Likewise, ACE has also been implemented together with FMM for the wideband analysis of electromagnetic 
phenomena \cite{vikram09}, with analytically derived error bounds that have been demonstrated via numerical experimentation.

\subsection{Earlier Work in Periodic Tree-Based Methods}

While tree-based methods (including ACE) have been studied extensively in the context of non-periodic problems \cite{nishimura02,vikram_review09}, their adaptation to periodic problems is encountered less frequently in the 
available literature. We attribute this to two factors, (i) periodic boundary conditions are often employed in situations in which they are already being used to decrease the effort required for a particular calculation, and 
(ii) a difficulty in constructing and evaluating the necessary translation operators.  This first factor essentially implies that tree-based methods will only be useful for periodic problems in which the unit cell is either very 
large, or very densely discretized.  This is, however, problem-dependent, and a number of interesting applications exist in which these requirements are met.  The latter factor stems from  the nature of the periodic Green's 
function for long-range interactions, namely that it is typically some manner of infinite sum, in which case the translation operator will not only be difficult to derive, but might require significant computational overhead. 

Early work in adapting tree-based methods to periodic systems were focused on Coulombic systems. In their seminal paper on the FMM \cite{greengard87}, Greengard and Rokhlin implement periodic boundary conditions, as well as 
Dirichlet and Neumann, for the two-dimensional Coulomb potential.  Schmidt and Lee later extended this approach by incorporating rapidly convergent Ewald summations \cite{schmidt91,schmidt97}.  Challacombe, {\em et. al.} 
\cite{challacombe97} published results based upon the efficient and accurate evaluation of lattice sums of spherical harmonics, enabling improvements in both the computational cost and memory overhead, relative to extant methods 
of the time.  One particularly interesting extension due to Lambert, et al \cite{lambert96} utilizes the hierarchical structure of the FMM to accelerate the aggregation of multipole expansions of periodic image cells, avoiding 
the direct evaluation of lattice sums.  Similar methods for the periodic Helmholtz kernel have seen more punctuated development.  While Rokhlin and Wandzura have presented work applying the FMM to periodic Helmholtz problems 
\cite{rokhlin94}, the extent of this work was limited to the rapid calculation of matrix elements arising in the solution of electromagnetic integral equations, rather than the  calculation of the potential itself (i.e. the 
associated matrix-vector product).  The FMM was not successfully employed in the iterative solution of electromagnetic integral equations until over a decade later in the work of Otani and Nishimura \cite{otani08a}.

The majority of the previously referenced tree-based methods incorporate periodic boundary conditions only at the top of the tree, corresponding to a multipole expansion of the entire unit cell.  The local expansion due to 
the influence of the rest of the lattice, excepting the nearfield of the unit cell, is evaluated using lattice sums.  At all levels below the top, free space translation operators are used, taking into account not only the 
effect of boxes lying inside of the unit cell, but those in its nearfield as well.  In this paper, we follow a more conventional approach, viz., use the addition theorem for the full periodic Green's function for 
multipole-to-local translations.  This is very similar in spirit to what has been done using interpolatory methods \cite{li10conf,li10,shi10}, and in fast time domain methods \cite{chen05} in electromagnetics. 
As a consequence, all such translations will be restricted to the interior of a single unit cell, reducing the total number of operations required per tree traversal relative to these other methods, at the expense of 
requiring more complex translation operators.  Given differences between test architectures and implementation, it is difficult to draw direct comparisons between timings for our method and others available in the 
literature.  However, we provide extensive results in Section \ref{results_section} that demonstrate an exceptional acceleration relative to direct evaluation, and breakeven points that are clearly very competitive with extant methods.

\subsection{Outline of Contents}

In this paper, we demonstrate the extension of the ACE algorithm to a wide array of periodic potentials, from Coulomb to Yukawa to Helmholtz.  In doing so, we discuss the algorithmic changes required, namely 
the derivation and evaluation of periodic translation operators and how interaction lists are constructed.  We do not seek to tie this work to the solution of 
any particular problem (i.e. integral equation solvers, $N$-body dynamics, etc), although we note that we have adapted our method to the solution of 
integral equations which arise in the analysis of electromagnetic wave propagation, and have submitted it to a more appropriate forum \cite{baczewski11}.  The principal contributions of this paper are as follows:
\begin{enumerate}
 \item Derivation of the necessary translation operators for periodic Helmhholtz, Yukawa, and Coulomb potentials on physically relevant lattices (singly, doubly, and triply periodic)
 \item Algorithmic changes for constructing ACE interaction lists on periodic domains
 \item Error bounds on the associated expansions 
\end{enumerate}
In Section \ref{math_framework}, we provide mathematical details concerning the class
of problems we aim to solve.  Our approach is sufficiently general that we can succinctly present details for periodic Coulomb, Yukawa, and Helmholtz potentials for singly, doubly, and triply periodic lattices.  In 
Section \ref{ace_framework}, the ACE algorithm is reviewed, and details of its implementation for periodic domains are provided.  Finally, in Section \ref{results_section}, error convergence and scaling are 
demonstrated.  Details concerning periodic Green's functions, derivations of the necessary ACE translation operators, and the associated error bounds are given in the Appendices. This is done to improve the overall 
readability of the manuscript.

\section{Mathematical Framework}\label{math_framework}

\subsection{Statement of the Problem}
\label{statement}

Consider a domain, $\Omega \subset \RRR$, containing a source distribution, $\rho(\br)$, that gives rise to an unknown potential, $\psi(\br)$, governed by one of 
the following partial differential equations (PDEs):
\begin{subequations}
\label{pdes}
\begin{align}
 \left(\nabla^2 + \kappa^2\right) \psi(\br) &= -\rho(\br)~~~\text{(Helmholtz~Equation)}\\
 \left(\nabla^2 - \kappa^2\right) \psi(\br) &= -\rho(\br)~~~\text{(Yukawa~Equation)}\\
 \nabla^2 \psi(\br) &= -\rho(\br)~~~\text{(Poisson~Equation)}
\end{align}
\end{subequations} 
Here, $\kappa \in \mathbb{R}^{+}$, is a problem-dependent constant, which corresponds to the wavenumber for the Helmholtz equation, and an inverse screening length 
for the Yukawa equation.  Assume that $\rho(\br)$ is periodic with respect to a $\mu$-dimensional lattice, $\Lmu$, defined as:
\begin{equation}
 \Lmu := \left\lbrace \bt(\bn_\mu) = \sum_{i=1}^{\mu} n_i \lv_i | n_i \in \mathbb{Z} \right\rbrace
\end{equation}
Here, $\bn_\mu$ is used as a short-hand notation for the $\mu$-tuple ($n_1,\ldots,n_\mu$), and $\lv_i$ are the primitive vectors associated with some Bravais lattice.  For the 
sake of simplicity, we will consider a simple orthorhombic lattice.  One can define the associated reciprocal lattice, $\Lmu^*$:
\begin{equation}
 \Lmu^* := \left\lbrace \displaystyle \bk(\bn_\mu) = \sum_{i=1}^{\mu} n_i \rv_i | n_i \in \mathbb{Z} \right\rbrace
\end{equation}
Here, $\rv_i$ are the primitive reciprocal lattice vectors, which together with the primitive lattice vectors satisfy $\lv_i \otimes \rv_j = 2\pi \bKron_{ij}$, where 
$\bKron_{ij}$ is the Kronecker tensor. \\

\begin{figure}[ht]
    \centering
    \includegraphics[scale=0.5]{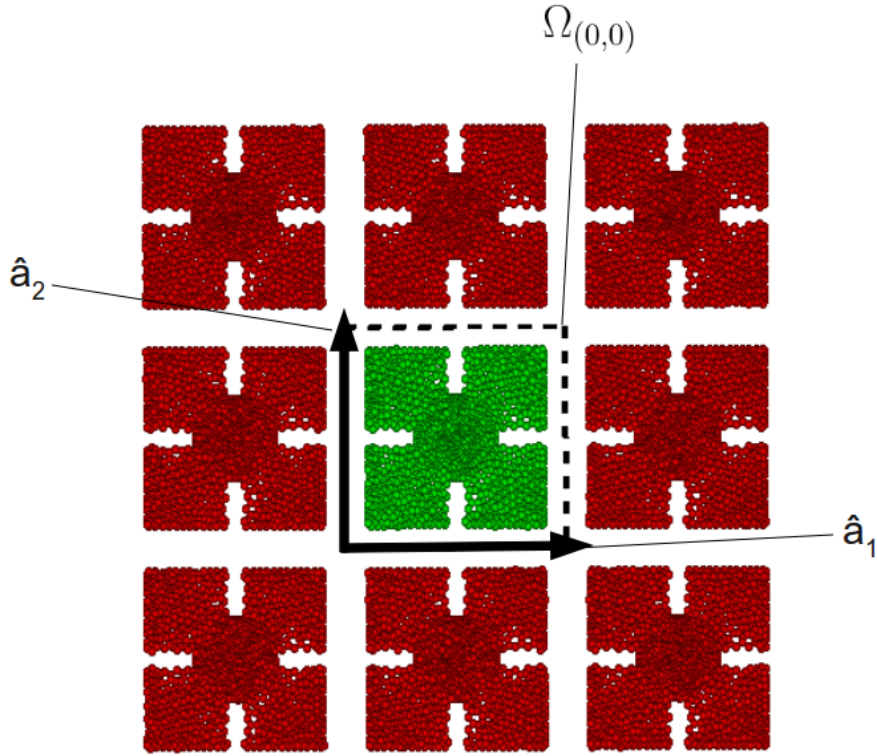}
    \caption{An exemplary periodic domain on a $2$-dimensional lattice.  The central primitive cell and its nearest image cells are illustrated.  Spheres are intended
    to represent $\rho(\br)$ in such a way that its periodicity is evident. \label{pgeom}}
\end{figure}

Using these definitions, we can describe the periodicity of $\rho(\br)$ as follows:
\begin{equation}
 \label{pbc_rho}
 \rho(\br + \bt(\bn_\mu)) = \rho(\br):\bt(\bn_\mu) \in \Lmu
\end{equation}
Given this constraint on $\rho(\br)$, we can fully characterize it over a reduced volume, namely the primitive cell, $\Omega_{\mu} \subset \Omega$, the minimal
set which reflects the translational symmetry of the lattice.  We can completely reconstruct $\Omega$ from a union of primitive cells shifted by all lattice vectors
in $\Lmu$:
\begin{equation}
\Omega = \left\lbrace \Omega_\mu + \bt(\bn_\mu) | \bt(\bn_\mu) \in \Lmu \right\rbrace 
\end{equation}
Here, we refer to the set $\Omega_{\mu} + \bt(\bn_\mu) = \Omega_{\bn_{\mu}}$ as the $\bn_\mu$th image cell, where $\Omega_{(0,..,0)} = \Omega_\mu$ is the central
primitive cell.  We denote the measure of space occupied by a single primitive cell in the subspace of the lattice as $\uA_\mu$ (i.e. $\uA_1=$length of primitive cell, etc).
Fig. (\ref{pgeom}) provides a pictorial representation of $\Omega$ for $\mu=2$. \\

By the translational invariance of the PDEs in (\ref{pdes}), that $\rho(\br)$ is periodic is sufficient to guarantee the periodicity of $\psi(\br)$, as well:
\begin{equation}
 \label{pbc_psi}
 \psi(\br) = \psi(\br + \bt(\bn_\mu)):\bt(\bn_\mu) \in \Lmu
\end{equation}
While this periodicity will suffice to constrain all boundary conditions when the codimension of the lattice is zero, for situations in which it is non-zero we must consider
the $3-\mu$ unspecified boundaries.  We consider these boundaries to behave as unbounded space, as this is frequently the physically relevant choice for a number of modeling 
scenarios.  In other words, for the Yukawa and Coulomb potentials, we consider the situation in which the potential decays as we recede away from the lattice, and for the 
Helmholtz potential, we consider a Sommerfeld boundary condition.  For the Helmholtz equation, we may also be interested in quasi-periodic boundary conditions, wherein 
a phase factor due to a non-zero Floquet wavenumber will arise as each cell is traversed.  However, in the interest of maintaining a unified approach to the three potentials, we 
relegate a full discussion of this to Appendix \ref{quasi_appendix}.\\

Having fully specified the problem, i.e., PDE and boundary conditions, we seek solutions for $\psi(\br)$.  Given the periodic boundary conditions, we can determine our 
potential solution completely in terms of $\rho(\br)$ for $\br \in \Omega_\mu$:
\begin{equation}
 \label{potential_eqn}
 \psi(\br) = \int \limits_{\Omega_\mu} d\br' G_{\mu}(|\br-\br'|) \rho(\br') 
\end{equation}
Here, $G_{\mu}(|\br-\br'|)$ is the appropriate periodic Green's function; a more complete and rigorous discussion of its derivation and evaluation can be found in Appendices \ref{gf_appendix} and \ref{gf_calc}.  Given its rapid and absolute convergence in and away from $\Lmu$, we utilize the Ewald representation of the periodic
Green's function \cite{ewald21,deleeuw79,deleeuw80,salin00,capolino07,jordan86,lovat08}:
\begin{equation}
\label{ewald_sum}
 G_\mu(|\br-\br'|) = \displaystyle\sum_{\bt(\bn_\mu)} \Ewr(|\br-\br'+\bt(\bn_\mu)|) + 
                     \displaystyle\sum_{\bk(\bn_\mu)} \Ewk(\br-\br',\bk(\bn_\mu))
\end{equation}
Following the usual convention, the first summation will be referred to as the `real sum' and the second as the `reciprocal sum'.  The functional form of the terms in the real sum depend upon the type of potential being evaluated:
\begin{subequations}
\label{real_ewald}
\begin{align}
 \Ewr(|\br-\br'+\bt(\bn_\mu)|)&=\displaystyle\sum_{\pm} \frac{e^{\pm i\kappa |\br-\br'+\bt(\bn_\mu)|}}{8\pi |\br-\br'+\bt(\bn_\mu)|} erfc\left(\eta |\br-\br'+\bt(\bn_\mu)| \pm i\frac{\kappa}{2\eta} \right)~~~&\text{(Helmholtz)} \\
                              &=\displaystyle\sum_{\pm} \frac{e^{\pm \kappa |\br-\br'+\bt(\bn_\mu)|}}{8\pi |\br-\br'+\bt(\bn_\mu)|} erfc\left(\eta |\br-\br'+\bt(\bn_\mu)| \pm \frac{\kappa}{2\eta} \right)~~~&\text{(Yukawa)} \\
                              &=\frac{1}{4\pi |\br-\br'+\bt(\bn_\mu)|} erfc\left(\eta|\br-\br'+\bt(\bn_\mu)|\right)~~~&\text{(Poisson)}
\end{align}
\end{subequations}
Here, $erfc$ is the complimentary error function \cite{abramowitz}, and $\eta \in \mathbb{R}^+$ is deemed the splitting parameter.  The form of the terms in the reciprocal sum depend upon both $\mu$ and the type of potential, 
by way of a function $\alpha(\bn_\mu)$:
\begin{subequations}
\label{recip_ewald}
\begin{align}
  \Ewk(\br-\br',\bk(\bn_\mu))&=\frac{e^{i\bk(\bn_1)\cdot \br_l}}{4\pi \uA_1} \displaystyle\sum_{\mu=0}^{\infty} \frac{(-1)^\mu}{\mu!} (|\br_t|\eta)^{2\mu} E_{1+\mu}\left(\frac{\alpha^2(\bn_1)}{4\eta^2}\right)~~~&\text{($\mu=1$)} \\
                             &=\frac{e^{i\bk(\bn_2)\cdot \br_l}}{4\uA_2 \alpha(\bn_2)} \displaystyle\sum \limits_{\pm} e^{\pm \alpha(\bn_2)|\br_t|} erfc\left(\frac{\alpha(\bn_2)}{2\eta} \pm \eta|\br_t| \right)~~~&\text{($\mu=2$)} \\
                             &=\frac{e^{i\bk(\bn_3)\cdot\br_l-\alpha^2(\bn_3)^2/4\eta^2}}{\uA_3 \alpha^2(\bn_3)}~~~&\text{($\mu=3$)}
\end{align}
\end{subequations}
Here, $E_n$ is the exponential integral of $n$th order \cite{abramowitz}, and $\br_l$ and $\br_t$ are projections of $\br-\br'$ which are parallel and transverse to the span of the lattice vectors, respectively, and 
$\alpha(\bn_\mu)$ is defined as follows for the potentials of interest:
\begin{subequations}
\begin{align}
 \label{spectral_param}
 &\alpha(\bn_\mu) = \sqrt{\left|\bk(\bn_\mu)\right|^2 - \kappa^2}~~&\text{(Helmholtz)} \\
 &\alpha(\bn_\mu) = \sqrt{\left|\bk(\bn_\mu)\right|^2 + \kappa^2}~~&\text{(Yukawa)} \\
 &\alpha(\bn_\mu) = \left| \bk(\bn_\mu) \right|~~&\text{(Poisson)}
\end{align}
\end{subequations}
It is evident that this form of the Green's function possesses some singularities that will complicate the evaluation of Eqn. (\ref{potential_eqn}), namely when 
$|\br-\br'| \to 0$ or $\alpha(\bn_\mu) \to 0$.  Unless otherwise indicated, we will implicitly exclude these singular contributions to the potential, and leave a more complete
discussion of their proper treatment to Appendix \ref{gf_appendix}. \\

Having specified $G_\mu(|\br-\br'|)$, given $\rho(\br)$,  $\psi(\br)$ can be calculated for $\br \in \Omega_\mu$ using Eqn. (\ref{potential_eqn}), furnishing a solution to the 
PDEs in Eqn. (\ref{pdes}) subject to appropriate boundary conditions.  The primary focus of this paper will be the rapid evaluation of a discrete form of the convolution in (\ref{potential_eqn})
using the ACE algorithm.  In what follows, we will briefly discuss this discretization, and then move onto the details of the ACE algorithm.
  
\subsection{Discretization of the Problem}

In discretizing Eqn. (\ref{potential_eqn}), without loss of generality, we consider the case where $\rho(\br)$ can be expressed as $N$ point sources distributed 
throughout $\Omega_\mu:$
\begin{equation}
 \rho(\br) = \displaystyle\sum_{\beta=1}^{N} q_\beta \delta(\br-\br_\beta)
\end{equation}  
Here, $\br_\beta \in \Omega_\mu$ is the location of the $\beta$th discrete source, and $q_\beta$ is its associated weight.  We can now write $\psi(\br)$ as:
\begin{equation}
 \psi(\br) = \displaystyle\sum_{\beta=1}^{N} q_\beta G_{\mu}(|\br-\br_\beta|) 
\end{equation}
Without loss of generality, we consider the calculation of the mutual interaction between these sources, viz. Eqn. (\ref{potential_eqn}) evaluated at all source points.  
We can reformulate this calculation as a matrix equation as follows:
\begin{subequations} 
 \begin{align}  
  \psi(\br_\alpha) = \sum \limits_{\alpha \neq \beta} q_\beta G_\mu(|\br_\alpha-\br_\beta|) ~~&\rightarrow~~ \bPsi_{\alpha} = \left(1 - \bKron_{\alpha \beta}\right)G_\mu(|\br_\alpha-\br_\beta|) \bQ_\beta \\
  \bPsi_{\alpha} &= \bG_{\alpha \beta} \bQ_\beta \label{mat-vec}
 \end{align}
\end{subequations}
Here, $\bG$ is an $N \times N$ matrix in which the self-interaction terms are explicitly excluded and $\bQ$ and $\bPsi$ are $N \times 1$ column vectors with entries $\bQ_{\beta} = q_\beta$ 
and $\bPsi_{\alpha} = \psi(\br_\alpha)$.  From this form of the potential, it is evident that the calculation of the potential will require $\mathcal{O}(N^2)$ operations and $\mathcal{O}(N^2)$ storage 
using direct methods.  It is the goal of this work to demonstrate that using the ACE algorithm, the total cost of potential computation is reduced to $\order(N)$ in time and $\order(N)$ in storage.  \\

Thus far, we have presented the convolution in Eqns. (\ref{potential_eqn}) and (\ref{mat-vec}) as the solution to a particular set of PDEs.  It is important to note that this 
does not limit the scope of this work to problems in which the explicit form of $\rho(\br)$ in $\Omega_\mu$ is known a priori.  In the integral equation formulation of 
numerous problems in applied math/physics, $\rho(\br)$ (discretely, $\bQ$) is an unknown to be resolved, and $\psi(\br)$ ($\bPsi$) is known.  In this context, iterative 
methods for the solution of Eqn. (\ref{mat-vec}) will require the repeated evaluation of matrix-vector multiplication with $\bG$. To this end the utility of fast potential 
evaluators, including ACE, is well-documented for a multitude of problems \cite{otani08a,otani09,vikram07,vikram09,vikram_review09,otani08b,chew97,lu06}.

\section{Rapid Evaluation of Periodic Potentials}\label{ace_framework}

\subsection{Description of the Algorithm} 

ACE and other FMM-type methods achieve an $\order(N)$ cost in timing and storage by approximating Eqn. (\ref{mat-vec}) as follows:
\begin{equation}
 \label{ace}
 \bPsi_\alpha = \bG_{\alpha \beta} \bQ_{\beta} \approx \bG_{\alpha \beta}^{near} \bQ_{\beta} + \mathcal{L}^{ACE}(\bQ_\beta)
\end{equation}
Here, $\bG^{near}_{\alpha \beta}$ is a sparse matrix, carrying only entries of $\bG_{\alpha \beta}$ describing interactions between source-observer pairs which are in some metric, `near', and 
$\mathcal{L}^{ACE}$ is some composition of linear operators which approximates the interactions between the remaining `far' source-observer pairs. This operator will be defined more explicitly 
in Section \ref{tree_trav}.  The ACE algorithm describes the metric that demarcates `near' and `far' interactions based upon a hierarchical decomposition of $\Omega_\mu$, and then provides rules for performing 
both operations in Eqn. (\ref{ace}) in a manner that requires $\order(N)$ operations, and $\order(N)$ storage.  \\

Such a hierarchical decomposition is achieved by mapping all source and observer points onto a regular octree data structure, henceforth referred to as `the 
tree'.  This structure provides a natural metric by which `near' and `far' interactions can be separated, as well as a means by which the application of $\mathcal{L}^{ACE}$ 
can be mapped onto the traversal of the tree.  Using these notions, the ACE algorithm can be used to evaluate Eqn. (\ref{ace}) in the following steps: 
\begin{algorithm}
 \caption{ACE Algorithm}
 \begin{algorithmic}[1]
  \STATE Construct the tree based upon discretization of $\rho(\br)$ and $\psi(\br)$ in $\Omega_\mu$.
  \STATE Fill nearfield/farfield interaction lists based upon tree.
  \STATE Precompute nearfield matrix elements and ACE translation operators. 
  \STATE Compute $\bPsi_\alpha$ via sparse nearfield matrix multiplication and tree traversal. 
 \end{algorithmic}
\end{algorithm}

Steps 1-3 constitute pre-processing, which need be performed only once for a fixed source distribution.  Step 4 is the only recurrent 
step required for the repeated evaluation of a potential in which the entries of $\bQ$ vary.  In what follows, we describe each step in detail 
and provide mathematical substantiation in Section (\ref{tree_trav}).

\subsubsection{Constructing the Tree}
\label{tree_sect}

\vspace{0.1in} Tree construction is based upon the specification of the desired number of levels $N_l$, chosen to optimize the cost and/or error.  The primitive cell, 
$\Omega_\mu$ is recursively subdivided into boxes of equal volume $N_l-1$ times. A single level of the tree is defined by a set of boxes of equal volume, with 1 
being the level with the smallest (`leaf') boxes.  At a given level, a box subordinate to a larger box at the level above is deemed the child to the larger box's parent.  
Every box is assigned an address in octal, based upon the usual octree decomposition, allowing us to readily acquire the `genealogy' of a particular box given its address 
alone.  This method of addressing boxes is called Morton ordering or Z-space filling curves \cite{warren93}. Each box at the leaf level carries a list of the point sources/observers lying inside its boundaries.  Given the hierarchical structure of the tree, it is trivial 
to determine the point sources/observers subordinate to boxes at lower levels by recursively aggregating children until the leaf level is reached. \newline

\begin{figure}[ht]
    \centering
    \includegraphics[scale=0.2]{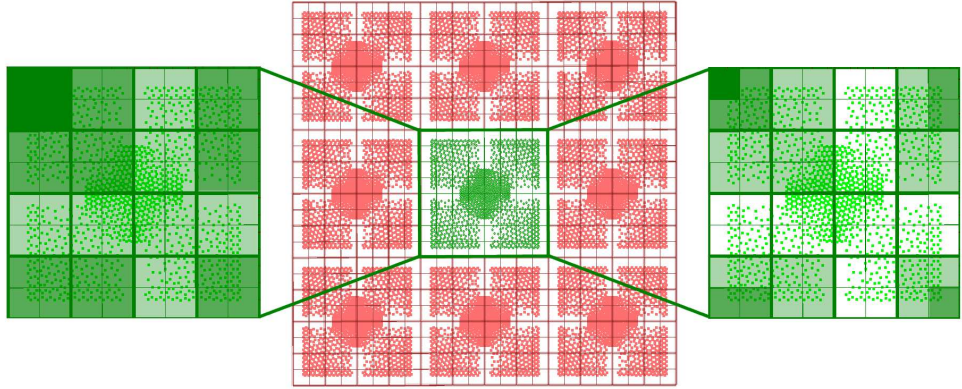}
    \caption{The `near' (green) and `far' (light green) interactions for source boxes (dark green) at different levels of the tree.  Sources and observers are co-located, with green points corresponding to sources 
defined in $\Omega_\mu$ (i.e. the primitive cell), and red points corresponding to their periodic images.  Red points/boxes are not actually stored, but need be considered in constructing interaction lists, as 
discussed in Section \ref{int_lists}.  \label{tree_geom}}
\end{figure}

As will be explicitly discussed in the next section, the construction of the interaction lists requires the consideration of not just boxes that lie inside of $\Omega_\mu$, 
but their nearest images as well. This is done to properly catalog near and far interactions, as is elucidated in the next section. The images of the cell are included by adding two 
fictitious levels together with the tree hierarchy that represents the nearest images of $\Omega_\mu$.  Figure \ref{tree_geom} illustrates this modification to the overall structure.
Note, these image boxes will not store any sources or observers, and are not 
involved in tree traversal.  Instead, they only serve as placeholders in addressing boxes, incurring a negligible computational overhead, and we do not consider their 
contribution to the height of the tree when it is referenced.  As the addressing scheme of a regular octree follows Morton ordering, boxes inside the primitive cell will 
fall within a contiguous address space \cite{warren93}.  This provides a simple means by which real boxes can be distinguished from image boxes.  As will be discussed in 
Section (\ref{int_lists}), knowledge of the nearest images of the primitive cell are essential in constructing the necessary interaction lists, which this extension of the 
conventional tree structure provides.  

\subsubsection{Filling Interaction Lists}
\label{int_lists}

\vspace{0.1in} The most crucial step in achieving a linear method using tree-based methods is choosing an appropriate rule for separating `near' and `far' interactions.
In non-periodic domains, this demarcation is a straightforward task.  Two boxes are considered to be `far' from one another if (i) they are separated by at least one box
length and (ii) their parents are not `far' from one another.  The first portion of this criterion is naturally tied to the distance between boxes, and ensures that spatial variations
in the relative potential of the two domains are limited.  The latter portion, on the other hand, ensures that `far' interactions are computed in $\order(N)$ operations. \newline

One of the primary challenges in adapting ACE to periodic domains is constructing a rule that meets both of these needs: minimal spatial variations in the relative potential as well as 
$\order(N)$ scaling.  This is somewhat non-trivial, in that the periodic boundary conditions map the problem onto a domain with a toroidal topology.  Consequently, the distance between
two boxes is no longer unique, as each box will possess images that effectively contribute to the relative potential through the periodic Green's function.  In determining whether or 
not two boxes in the primitive cell are `far' from one another, we must then construct a rule which gives consideration to all image boxes.  Such a rule is as follows:\newline
\begin{enumerate}
 \item The original boxes and {\bf all} of their images are separated by at least one box length. 
 \item Among the parents of both the original and image boxes, {\bf at least} one pair is not `far'.
\end{enumerate}

\vspace{0.1in} Fig. (\ref{tree_geom}) illustrates the `near' and `far' boxes for an exemplary leaf box in a three and four level tree.  The contents of boxes outlined in green are stored 
explicitly in the tree, whereas boxes outlined in red are part of the fictitious extra levels discussed in the previous Section.  For both the three and four level trees, it is worth noting 
the manner in which the nearfield wraps around the unit cell.  The source box residing in the upper left corner of the primitive cell will participate in nearfield interactions with 
observer boxes residing at each of the other 3 corners, in spite of their apparent distance.  We have observed poorer error convergence when using the usual, non-periodic rules for parsing 
interactions, so this is an important subtlety to keep in mind when adapting extant codes to periodic problems. \newline

Mechanically, the construction of the interaction lists is straightforward.  For a given level, we know the range of addresses which lie inside the primitive cell, so we can explicitly 
iterate over only these boxes.  This iteration begins at the $(N_l-1)$th level, just below the `root' of the real tree, at which all boxes are `near' each other.  At the next level down, we iterate over the 
children of these boxes, and apply the `near' versus `far' rule outlined above.  As we have addressed the nearest image boxes using Morton ordering, we can conveniently access not only the relative location 
of an image box given its corresponding real box, but its entire `genealogy'.  We continue to descend the tree, iterating over only real boxes at each level, until we have iterated over the leaf 
level, at which point the interaction lists are complete.  These lists are next employed in precomputation, in which they are used to construct a list of the necessary 
nearfield matrix elements and unique ACE translation operators that need be computed.

\subsubsection{Precomputation}

\vspace{0.1in} Precomputation can be broken into two stages, (i) nearfield matrix elements and (ii) ACE translation operators.  In constructing nearfield matrix elements, we iterate 
over all unique source-observer pairs and compute elements of $\bG^{near}$ directly.  Precomputation of ACE translation operators is slightly more complex.  All farfield interactions 
are sorted by the distance separating source and observer parent domains in each of the Cartesian coordinates.  Given the regularity that our decomposition of the domain imposes 
(i.e. all boxes at a given level have the same dimensions), and that the ACE translation operators only depend upon the relative separation of two boxes, and not their absolute 
position, there is significant degeneracy among the ACE translation operators.  Consequently, in sorting by domain separation, we can identify a subset of unique translation operators 
which can be computed once, stored, and recycled.  This is particularly important in periodic ACE, as the calculation of a single unique translation operator will possess
$\frac{(P+3)(P+2)(P+1)}{6}$ unique components, each of which is an Ewald-like infinite sum (see Eqn. \ref{trans_sum}), and thus non-trivial.  

\subsubsection{Tree Traversal}\label{tree_trav}

\vspace{0.1in} The evaluation of $\mathcal{L}^{ACE}(\bQ_\beta)$, in Eqn. (\ref{ace}), via tree traversal proceeds along the following five steps, common to any fast multipole-type method:
\begin{algorithm}[h]
 \caption{ACE Tree Traversal}
 \label{tree_alg}
  \begin{algorithmic}[1]
    \STATE {\bf Charge-to-Multipole (C2M):} Construct multipole expansions from point sources in each leaf box.
    \STATE {\bf Multipole-to-Multipole (M2M):} Aggregate multipole expansions at higher levels of the tree.
    \STATE {\bf Multipole-to-Local (M2L):} Translate multipole expansions about source domains to local expansions about observer domains.
    \STATE {\bf Local-to-Local (L2L):} Disaggregate local expansions at lower levels of the tree.
    \STATE {\bf Local-to-Observer (L2O):} Compute observer field from local expansions in each leaf box.
  \end{algorithmic}
\end{algorithm}

\hspace{-0.1in}In what follows, we provide some of the Theorems that describe Algorithm \ref{tree_alg} in a mathematically rigorous context.  First, however, as the ACE algorithm was developed in the 
language of Cartesian tensors, a brief overview of the necessary tensor notations is provided in exposition.  We denote a Cartesian tensor of rank $n$ by $\bA^{(n)}$.  In general, 
such a tensor consists of $3^{n}$ components from $\mathbb{C}$, indexed by the set $\lbrace\alpha_i~|~i\in\lbrace1,\ldots,n \rbrace, \alpha_i \in \lbrace 1,2,3 \rbrace \rbrace$, 
where an individual component is given as $\iA^{(n)}_{\alpha_1\ldots\alpha_n}$.  A totally symmetric tensor is one in which $\iA^{(n)}_{\alpha_1\ldots\alpha_n}$ is independent 
of any permutation on the indices, and can be represented by $(n+1)(n+2)/2$ independent components.  Consequently, we can index components of the totally symmetric tensor, $\bA^{(n)}$,
 as $\iA\left[n_1,n_2,n_3\right]$, where $n_i$ is the number of times that index $i$ occurs and $n_1+n_2+n_3=n$. An $n$-fold contraction between two tensors is denoted by $\cdot n \cdot $, 
such that $\bC^{(m-n)}=\bA^{(m)}\cdot n \cdot\bB^{(n)}$.  Finally, we denote an n-fold product of a rank 1 tensor, $\br$, with itself by $\br^{(n)}$.  
More details concerning Cartesian tensors can be found in References \cite{shanker07} and \cite{applequist83}. \newline

In what follows, we prescribe the computation of potentials that are  observed in a domain $\Omega_o \subset \Omega_\mu$, due to sources in a domain $\Omega_s \subset \Omega_\mu$, 
where the two domains are well-separated in the sense that they are in each others' farfield.  This framework implies that potentials in $\Omega_o$ due to other source clusters can be found using the same framework. Superordinate to these domains are their respective parent domains, $\Omega_s \subset \Omega_s^p$ 
and $\Omega_o \subset \Omega_o^p$.  The centroids of these domains are located at $\br_s$, $\br_s^p$, $\br_o$, and $\br_o^p$. Using these notations, Steps 1-5 of Algorithm \ref{tree_alg} are 
effected using the following five Theorems.  \\

We begin with a Theorem which provides a functional definition of a Cartesian multipole expansion at the leaf level.
\begin{thm}[Charge-to-Multipole Expansion (C2M)] \label{C2M}
The potential, $\psi(\br_\alpha)$, at any point $\br_\alpha \in \Omega_o$, due to $S$ sources at points $\br_i' \in \Omega_s$ with strength $q_i$ $(i \in \lbrace 1,2,\ldots,S \rbrace)$ can 
be expressed in terms of a Cartesian multipole expansion.
\begin{subequations}
\begin{align}
&\psi(\br) = \displaystyle\sum_{\beta=1}^{S} q_\beta G_\mu(|\br_\alpha-\br'_\beta|) = \displaystyle\sum_{n=0}^{\infty} \bM^{(n)} \cdot n \cdot \nabla^{(n)} G_{\mu}(\br_\alpha-\br_s), \\
&\bM^{(n)} = \displaystyle\sum_{\beta=1}^{S} (-1)^n \frac{q_\beta}{n!}(\br_\beta'-\br_s)^{(n)} \label{c2m_exp}
\end{align}
\end{subequations}
\end{thm}

\vspace{0.25in}We next consider the manner in which the origin of a multipole expansion can be shifted, in such a way that aggregate multipole expansions can be constructed at higher levels of 
the tree, based upon extant expansions at lower levels of the tree.
\begin{thm}[Multipole-to-Multipole Expansion (M2M)] \label{M2M}
A multipole expansion of $S$ sources about $\br_s$, $\bO^{(n)}$, can be expressed in terms of a multipole expansion about the point $\br_s^p$.
\begin{subequations}
\begin{align}
&\bM^{(n)} = \displaystyle\sum_{\beta=1}^{S} (-1)^n \frac{q_\beta}{n!}(\br_\beta'-\br_s^p)^{(n)} = \displaystyle\sum_{m=0}^{n}\displaystyle\sum_{P(m,n)} \frac{m!}{n!} (\br_s^p - \br_s)^{(n-m)} \bO^{(m)} \label{m2m_exp} \\
&\bO^{(n)} = \displaystyle\sum_{\beta=1}^{S} (-1)^n \frac{q_\beta}{n!}(\br_\beta'-\br_s)^{(n)}
\end{align}
\end{subequations}
\end{thm}
Here, it is important to note that Eqn. (\ref{m2m_exp}) is mathematically equivalent to Eqn. (\ref{c2m_exp}).  In other words, we do not incur additional error in shifting the origin of our multipole expansion, using 
(\ref{m2m_exp}) instead of (\ref{c2m_exp}).  It is this detail of the ACE algorithm which leads to an error which is completely independent of the height of the tree, a feature which is demonstrated to machine 
precision in \cite{shanker07,vikram10,vikram09}. \\

At some level in the tree, multipole expansions in the source domain are translated into local expansions in the observer domain.  The following Theorem expresses this process mathematically in terms 
of a translation operator.
\begin{thm}[Multipole-to-Local Translation (M2L)] \label{M2L}
For a multipole expansion about $\br_s^p$, $\bM^{(n)}$, a local expansion $\bL^{(n)}$ that produces the same field in $\Omega_o^p$ is given by:
\begin{subequations}
\begin{align}
&\psi(\br_\alpha) = \displaystyle \sum_{n=0}^{\infty} (\br_\alpha-\br_o^p)^{(n)} \cdot n \cdot \bL^{(n)} \\
&\bL^{(n)} = \displaystyle \sum_{m=n}^{\infty} \frac{1}{n!} \bM^{(m-n)} \cdot (m - n) \cdot \nabla^{(m)} G_\mu(|\br_o^p - \br_s^p|)
\end{align}
\end{subequations}
\end{thm}
%define tilde here!
Here, the translation operator is the set of all tensors, $\nabla^{(m)} G_\mu(|\br_o^p - \br_s^p|)$, where $m \in \mathbb{N}$.  The elements of these tensors correspond to coefficients of a Taylor series expansion of 
the Green's function about $|\br_o^p-\br_s^p|$.  For a rank $p$ tensor component of the translation operator, we write its individual components as:
\begin{equation}
 \nabla^{(p)} G_\mu(|\br_o^p-\br_s^p|)\left[p_x,p_y,p_z\right] = \partial_x^{p_x} \partial_y^{p_y} \partial_z^{p_z} G_\mu(|\br_o^p-\br_s^p|)~~:~~p_x + p_y + p_z = p
\end{equation}
As we are utilizing the Ewald representation for $G_\mu(|\br_o^p-\br_s^p|)$, we may apply the necessary partial derivatives to each term of Eqn. (\ref{ewald_sum}), yielding:
\begin{equation}
 \label{trans_sum}
 \nabla^{(p)} G_\mu(|\br_o^p-\br_s^p|)\left[p_x,p_y,p_z\right] = \displaystyle\sum_{\bt(\bn_\mu)} \partial_x^{p_x} \partial_y^{p_y} \partial_z^{p_z}\Ewr(|\br_o^p-\br_s^p+\bt(\bn_\mu)|) + 
                                                             \displaystyle\sum_{\bk(\bn_\mu)} \partial_x^{p_x} \partial_y^{p_y} \partial_z^{p_z}\Ewk(\br_o^p-\br_s^p,\bk(\bn_\mu))
\end{equation}
The elements of the translation operator arising from the real sum are given as:
\begin{subequations}
 \small
 \label{real_sum_expressions}
 \begin{align}  
  \nabla^{(p)} \Ewr(|\bR|)\left[p_x, p_y, p_z \right] &= \frac{(-1)^p}{4\pi^{3/2}|\bR|} \displaystyle\sum_{m=0}^{p} \sum_{\mu=0}^{\infty} C_{m}^{p_x,p_y,p_z} \frac{(\kappa^2 |\bR|^2/4)^{\mu}}{\mu!} \frac{\Gamma\left(\frac{p+m+1}{2}-\mu, \eta^2|\bR|^2 \right)}{|\bR|^{p+m}} &\text{(Helmholtz)}\\
					      &= \frac{(-1)^p}{4\pi^{3/2}|\bR|} \displaystyle\sum_{m=0}^{p} \sum_{\mu=0}^{\infty} C_{m}^{p_x,p_y,p_z} \frac{(-\kappa^2 |\bR|^2/4)^{\mu}}{\mu!} \frac{\Gamma\left(\frac{p+m+1}{2}-\mu, \eta^2|\bR|^2 \right)}{|\bR|^{p+m}} &\text{(Yukawa)} \\ 
		      \label{coulomb_real_ace}&= \frac{(-1)^p}{4\pi^{3/2}|\bR|} \displaystyle\sum_{m=0}^{p} C_{m}^{p_x,p_y,p_z} \frac{\Gamma\left(\frac{p+m+1}{2}, \eta^2|\bR|^2 \right)}{|\bR|^{p+m}} &\text{(Coulomb)}
 \end{align}
\end{subequations}
Here, $\Gamma(n,x)$ is the $n$th incomplete Gamma function and the coefficients, $C_{m}^{p_x,p_y,p_z} \in \mathbb{R}$.  A full derivation of this expression is provided in Appendix \ref{trans_ops}.  Expressions for 
the terms arising due to the reciprocal sum arise from straightforward partial differentiation of Eqn. (\ref{recip_ewald}), and are given as:
\begin{subequations}
 \small
\label{rec_sum_expressions}
 \begin{align}
    \nabla^{(p)} \Ewk(\bR,\bk(\bn_\mu))\left[p_x, p_y, p_z \right] &= (ik_x)^{p_x}\frac{e^{i\bk(\bn_1)\cdot \br_l}}{4\pi \uA_1} \displaystyle\sum_{\mu=0}^{\infty} \frac{(-\eta^2)^\mu}{\mu!} E_{1+\mu}\left(\frac{\alpha^2(\bn_1)}{4\eta^2}\right)\times \ldots \notag \\
       \ldots &\displaystyle \sum^{\mu}_{\substack{m=0 \\ 2\mu-2m-p_y \geq 0 \\ 2m-p_z \geq 0}} {\mu \choose m} \frac{(2\mu-2m)!(2m)!}{(2\mu-2m-py)!(2m-pz)!} R_y^{2\mu-2m-p_y} R_z^{2m-p_z} \hspace{0.45in} \text{($\mu=1$)} \label{top_k_1} \\
    &= (ik_x)^{p_x} (ik_y)^{p_y}\frac{e^{i\bk(\bn_2)\cdot \bR_l}}{4\uA_2 \alpha(\bn_2)} \displaystyle\sum \limits_{\pm} (\pm1)^{p_z} \left(\displaystyle\sum \limits_{m=1}^{p_z} {p_z \choose m} \frac{(-\eta)^m}{(\alpha(\bn_2))^{m-p_z}} \frac{2}{\sqrt{\pi}} H_{m-1}\left(\frac{\alpha(\bn_2)}{2\eta}\pm\eta R_z\right) \times \ldots \right. \notag \\ 
       \ldots &\left.e^{-\frac{\alpha(\bn_2)^2}{4\eta^2}-\eta^2 R_z^2} + (\alpha(\bn_2))^{p_z} e^{\pm \alpha(\bn_2)R_z} erfc\left(\frac{\alpha(\bn_2)}{2\eta}\pm\eta R_z\right)\right) \hspace{0.7in} \text{($\mu=2$)} \label{top_k_2} \\ \notag \\
    &= (ik_x)^{p_x} (ik_y^{p_y}) (ik_z^{p_z}) \frac{e^{i\bk(\bn_3)\cdot\bR_l-\alpha^2(\bn_3)/4\eta^2}}{\uA_3 \alpha^2(\bn_3)} \hspace{1.85in}\text{($\mu=3$)} \label{top_k_3}  \end{align}
\end{subequations}
Here, $R_i$ is the projection of $\bR$ along the $i$-axis, $E_{n}(x)$ is the $n$th exponential integral and $H_{n}(x)$ is the $n$th Hermite polynomial.  For $\mu=1$, we have assumed that the lattice lies along the x-axis, 
and for $\mu=2$, we have assumed that the lattice lies in xy-plane. \newline

As with multipole expansions, we can shift the origin of the local expansion in such a way that we can disaggregate local expansions centered about observer boxes at higher levels of the tree into 
expansions about observer boxes at lower levels.
\begin{thm}[Local-to-Local Expansion (L2L)] \label{L2L}
A local expansion $\bO^{(n)}$ centered about $\br_o^p$ can be expressed in terms of a local expansion about the point $\br_o$, using:
\begin{align}
\bL^{(n)} = \sum \limits_{m=n}^{\infty} {m \choose m-n} \bO^{(m)} \cdot (m-n) \cdot (\br_o-\br_o^p)^{(m-n)}
\end{align}
\end{thm}

\vspace{0.25in} Finally, we can compute the potential at a point $\br_\alpha$ in $\Omega_o$ from the local expansion centered about $\br_o$.
\begin{thm}[Local-to-Observer (L2O)] \label{L2O}
The potential, $\psi(\br_\alpha)$ can be expressed in terms of $\bL^{(n)}$, a local expansion centered about $\br_o$, using:
  \begin{align}
   \psi(\br_\alpha) = \sum \limits_{n=0}^{\infty} \bL^{(n)} \cdot n \cdot (\br_\alpha-\br_o)^{(n)}
  \end{align}
\end{thm}

\subsection{Theoretical Error Bounds}\label{bounds_section}

One of the primary advantages of fast multipole-type methods is the possibility of deriving mathematically rigorous bounds.  As discussed in \cite{shanker07}, there are two 
sources of error in the ACE algorithm: (i) $\varepsilon_m$ due to truncation of the Taylor expansion of multipoles at the level at which translation occurs and (ii) 
$\varepsilon_l$ due to truncation of the local expansion.  Both errors can be bounded for the potentials under consideration in this paper.  We begin by considering the 
error, $\varepsilon_m$, where the multipole expansion is truncated beyond $P$ harmonics.
\begin{subequations}
 \begin{align}
  \footnotesize
  \varepsilon_m &= \left|\psi(\br) - \displaystyle\sum_{n=0}^{P} \bM^{(n)} \cdot n \cdot \nabla^{(n)} G_\mu(|\br-\br^p_s|)\right| = \left|\displaystyle\sum_{n=P+1}^{\infty} \bM^{(n)} \cdot n \cdot \nabla^{(n)} G_\mu(|\br-\br^p_s|)\right| \\
             &= \left|\displaystyle\sum_{n=P+1}^{\infty} \bM^{(n)} \cdot n \cdot \nabla^{(n)} \left(\displaystyle\sum_{\bn_\mu} \left[\Ewr(|\br-\br^p_s+\bt(\bn_\mu)|) + \Ewk(|\br-\br^p_s|,\bk(\bn_\mu))\right]\right) \right|
 \end{align}
\end{subequations}
Invoking the absolute convergence of the Ewald representation, we can swap summation and differentiation, and then apply the triangle inequality:
\begin{subequations}
 \begin{align}
  \footnotesize
  \varepsilon_m &= \left|\displaystyle\sum_{n=P+1}^{\infty} \bM^{(n)} \cdot n \cdot \left(\displaystyle\sum_{\bn_\mu} \left[\nabla^{(n)} \Ewr(|\br-\br^p_s+\bt(\bn_\mu)|) + \nabla^{(n)} \Ewk(|\br-\br^p_s|,\bk(\bn_\mu))\right]\right) \right| \\  
  \label{m2l_error} \varepsilon_m &\leq \displaystyle\sum_{\bn_\mu} \left|\displaystyle\sum_{n=P+1}^{\infty} \bM^{(n)} \cdot n \cdot \nabla^{(n)} \Ewr(\bt(\bn_\mu),\ldots)\right| +
                   \displaystyle\sum_{\bn_\mu} \left|\displaystyle\sum_{n=P+1}^{\infty} \bM^{(n)} \cdot n \cdot \nabla^{(n)} \Ewk(\bk(\bn_\mu),\ldots)\right| 
 \end{align}
\end{subequations}
Here, we have separated $\varepsilon_m$ into separate contributions from the real and reciprocal sums.  
\begin{equation}
 \varepsilon_m \leq \displaystyle\sum_{\bn_\mu} \varepsilon_{m,r}(\bn_\mu) + \displaystyle\sum_{\bn_\mu} \varepsilon_{m,k}(\bn_\mu)
\end{equation}
Using similar arguments, we can arrive at an analogous expression for $\varepsilon_l$:
\begin{subequations}
 \begin{align}
  \varepsilon_l \leq & \displaystyle\sum_{\bn_\mu} \left|\displaystyle\sum_{n=P+1}^{\infty} (\br-\br_o)^{(n)} \cdot n \cdot \displaystyle\sum_{m=n}^{\infty} \bM^{(m-n)} \cdot (m-n) \cdot \nabla^{(m)} \Ewr(\bt(\bn_\mu),\ldots)\right| + \ldots \notag \\
                  & \displaystyle\sum_{\bn_\mu} \left|\displaystyle\sum_{n=P+1}^{\infty} (\br-\br_o)^{(n)} \cdot n \cdot \displaystyle\sum_{m=n}^{\infty} \bM^{(m-n)} \cdot (m-n) \cdot \nabla^{(m)} \Ewk(\bk(\bn_\mu),\ldots)\right| \\
  &\hspace{1in} \varepsilon_l \leq \displaystyle\sum_{\bn_\mu} \varepsilon_{l,r}(\bn_\mu) + \displaystyle\sum_{\bn_\mu} \varepsilon_{l,k}(\bn_\mu)
 \end{align}
\end{subequations}
In both cases, it is evident that the total error is bounded by the sum of errors incurred in reconstructing each term of the respective sums as a Taylor expansion. \\

Given this form for the bound on the total error, we can derive bounds on $\varepsilon_{m,r}(\bn_\mu)$ and $\varepsilon_{l,r}(\bn_\mu)$ based purely upon the potential being 
evaluated, as the codimensional dependency is implicit.  Similarly, $\varepsilon_{m,k}(\bn_\mu)$ and $\varepsilon_{l,k}(\bn_\mu)$ can be bounded purely based upon the codimension 
under consideration, as the dependency on the type of potential is implicit.  We present exemplary expressions for bounds on terms in the Coulombic real sum for arbitrary 
codimension, and for terms in the $\mu=3$ reciprocal sum for an arbitrary potential.  The bound on Coulombic terms is given as:
\begin{equation}
 \label{bound1}
 \varepsilon_{m,r}(\bn_\mu) \leq C \frac{a^{P+1}}{4\pi^{3/2} (P+1)! |\bR|} \left|(P+1)\frac{\Gamma(P+3/2,(1-a)\eta^2|\bR|^2)}{(1-a)^{P+3/2}} - a\frac{\Gamma(P+5/2,(1-a)\eta^2|\bR|^2)}{(1-a)^{P+5/2}} \right|
\end{equation}
Here, $a=|\br_{i,max}|/|\bR|<1$, where $|\bR|=|\br-\br_s+\bt(\bn_\mu)|$, $|\br_{i,max}|$ is the distance from the source box center to its furthest point source, and 
$C \in \mathbb{R}^{+}$.  A detailed derivation of this bound is provided in Appendix \ref{error_appendix}.  A similar procedure can be used to derive bounds on the 
Helmholtz and Yukawa potentials.  The bound on the $\mu=3$ reciprocal terms is given as:
\begin{equation}
 \label{bound2}
 \varepsilon_{m,k}(\bn_\mu) \leq C \frac{(\bk(\bn_3)\cdot\br_{i,max})^{P+1}}{(P+1)!} \left|\frac{e^{-\alpha^2(\bn_3)/4\eta^2}}{\uA_3 \alpha^2(\bn_3)} \right|
\end{equation}
While the proof is straightforward, it is also furnished in Appendix \ref{error_appendix}.  In both cases, the bound on the terms contributing to $\varepsilon_l$ can be derived 
following a procedure similar to that outlined in Appendix \ref{error_appendix}, as well. \\

For physically relevant parameters, we find that the dominant contributions to the error are due to terms in the sum on the interval $0 \leq |\bn_\mu| \leq 2$, 
as one may intuitively expect on the basis of the rapid convergence of the Ewald sum.  Further, we can see that as $P \to \infty$, the error in each term can be 
made arbitrarily small. \\

While bounds can be derived for the other potentials and codimensions under consideration, they are considerably looser, and thus ommited.  One of the primary reasons 
for this looseness is the anisotropy with which we expect our expansions to exhibit in their convergence.  This expectation is based upon two sources of anistropy, 
the discrete rotational invariance of the lattice, and variations in the behavior of the Green's function in and out of the lattice.  The former source will be prevalent 
in all of the potentials considered in this work, whereas the latter is limited to situations in which the codimension of the lattice is non-zero.  This stands in contrast 
to the problems to which ACE has previously been applied, wherein the kernels of the associated potentials are spherically symmetric, and error bounds are isotropic.  
In Section \ref{results_section}, we present numerical error convergence data to provide a more practical perspective on the accuracy of our expansions.    

\subsection{Computational Complexity}\label{cost}

The computational cost of the ACE algorithm has been previously analyzed in detail for non-periodic problems \cite{shanker07}.  Here, we provide a brief review of the dominant costs and highlight minor differences which arise in 
adapting ACE to periodic systems.  In doing so, we consider a primitive cell in which $N$ co-located source/observer points are randomly distributed.  These points are mapped onto an $N_l$ level tree, where the number of boxes 
at level $l$ is $B_{l}$ and $B_{l-1} = 8 B_{l}$.  The average number of unknowns per leaf box is denoted $s$, i.e. $N/s = B_{1}$, and the total number of boxes at all levels $\sum_{l=1}^{N_l} B_{l} \sim \frac{N}{s} \sim B_{1}$.  
The total cost of evaluating Eqn. (\ref{ace}), truncating all expansions at $P$th order, can be broken down into nearfield (NF), C2M, M2M, M2L, L2L, and L2O costs, each of which is summarized below:\\

\begin{enumerate}
\small
 \item {\bf NF:} $s^2$ operations per `nearfield' neighbor $\times$ 27 `nearfield' neighbors per leaf box $\times$ $B_{1}$ leaf boxes: $C_{NF} = 27N s$
 \item {\bf C2M:} $s$ operations per component $\times$ $\sim \frac{P^3}{6}$ unique multipole components $\times$ $B_{1}$ leaf boxes: $C_{C2M} = \frac{N}{6} P^3$
 \item {\bf M2M:} $\sim \frac{P^6}{720}$ operations per M2M translation $\times$ $\sim B_{1}$ M2M translations: $C_{M2M} = \frac{N}{s} \frac{P^6}{720}$
 \item {\bf M2L:} $\sim \frac{P^6}{720}$ operations per `farfield' neighbor $\times$ $\sim$ 56 `farfield' neighbors per box $\times$ $\sim B_{1}$ boxes: $C_{M2L} = 56 \frac{N}{s} \frac{P^6}{720}$
 \item {\bf L2L:} Same as M2M: $C_{L2L} = \frac{N}{s} \frac{P^6}{720}$
 \item {\bf L2O:} Same as C2M: $C_{L2O} = \frac{N}{6} P^3$
\end{enumerate}
This leaves us with the total cost:
\begin{equation}
 C_{tot} = N \left(27 s + \frac{P^3}{3} + 58\frac{P^6}{720 s}\right) \sim \mathcal{O}(N)
\end{equation}
The multiplicative factor behind $N$ is minimized if we choose a density of $s \approx \frac{P^3}{18}$.  As discussed in Section (\ref{tree_sect}), rather than 
specifying a minimum box size to achieve this ideal density of unknowns prior to constructing the tree, we specify the number of levels, $N_l$.  Knowing that 
$B_{1}=8^{N_l-1}$, this ideal density of unknowns is then realized for $N_l = \left[1 + \log_8(\frac{18N}{P^3})\right]$.

Excepting an improvement that we have made in our algorithm since the publication of \cite{shanker07}, this is essentially the same cost structure.  In our 
original publication, the number of `farfield' neighbors per box is given as 189, here we have reduced it to 56.  This is achieved by taking advantage of 
the exact up/down tree traversal operations in ACE.  In situations in which a box is in the `farfield' of all of the children of a given parent, the M2L translation 
will occur at the level of the parent.  For a given box, there will be $27-8=19$ such parent level interactions, and $64-27=37$ child level interactions, leading 
to 56 such interactions in aggregate.  This change in the way `farfield' interactions are treated is independent of whether ACE is being used for periodic 
or non-periodic problems. \\

The only significant difference in cost between periodic and non-periodic ACE is in the prefactors in the C2M and M2L costs.  The 27 `nearfield' neighbors per leaf box 
and 56 `farfield' neighbors per box are upper bounds, and in the case of non-periodic ACE, only realized for boxes that have no faces touching the boundary of 
the computational domain.  For boxes on the boundary, however, both the number of near and far interactions will be reduced, the net effect of which is a 
practically negligible change in the optimal value for $s$ and the overall prefactor.  In periodic ACE, however, boundary boxes essentially participate in 
near and far interactions as if they were on the interior of the computational domain, given our revised rule for filling interaction lists.  In this sense, 
periodic ACE is actually closer to the idealized cost given above.

\section{Results}\label{results_section}
In this Section, we present the results of numerical experiments carried out using the ACE algorithm.  These results are intended to validate the utility of our method as a fast and accurate means of evaluating 
the periodic potentials discussed in this paper.  Error convergence tests were performed on a desktop computer with a dual core Intel Pentium D clocked at 3.20GHz with 3GB RAM, running Linux OS.  Scaling tests were performed using 
a single node at the Michigan State University High Performance Computing Center (HPCC).  Each node is equipped with two quad core Intel Xeons clocked at 2.93GHz with access to 47GB RAM.  All code was compiled using the Intel 
Fortran Compiler.  We have not exploited any degree of parallelism in generating these results; HPC resources were employed to raise the ceiling on the number of unknowns for our scaling tests.

\subsection{Error Convergence}
In demonstrating error convergence, we consider only the contributions to the total potential arising due to farfield interactions. $\bPsi^{ACE}_{\alpha} = \mathcal{L}^{ACE}(\bQ_\beta)$ is computed using the ACE algorithm whereas 
$\bPsi^{direct}_{\alpha} = \bG_{\alpha \beta} \bQ_{\beta} - \bG_{\alpha \beta}^{near} \bQ_{\beta}$ is computed directly.  We do not explicitly compute this difference, but simply ignore nearfield pairs in evaluating 
the potential.  The relative error in the $L_2$-norm is reported in all numerical experiments:
\begin{equation}
 \varepsilon_{far} = \sqrt{\frac{|\bPsi^{ACE}_{\alpha}-\bPsi^{direct}_{\alpha}|^2}{|\bPsi^{direct}_{\alpha}|^2}}
\end{equation}
As farfield contributions will tend to be slightly smaller than those due to nearfield interactions, this is in some sense a `worst-case' metric for error, and one can typically expect an order of 
magnitude improvement in the error for the entire potential. \newline

We first demonstrate that we can achieve arbitrary precision using the ACE algorithm by increasing the order above which our expansions are truncated, $P$.  Results are presented for each of the 9 permutations of 
potentials/codimensions; all computations are done for a random distribution of co-located source/observer points.  In all tests, the locations of 1000 sources were chosen from a random distribution on a line ($\mu=1$), plane 
($\mu=2$), or cube ($\mu=3$). The magnitude of the source was chosen at random from a uniform distribution on the interval $[0,1)$.  All Ewald sums were evaluated to a relative error of $\varepsilon_{GF}=10^{-5}$.  The error, $\varepsilon_{far}$ is calculated under these conditions as $P$ is varied from 1 to 11, and the results for $\mu=1,2,3$ are given in Tables \ref{1d_p_convergence}, \ref{2d_p_convergence}, and \ref{3d_p_convergence}, 
respectively, for all 3 potentials.  From these results, it is evident that the error in approximating all 9 potentials with ACE expansions decreases uniformly as $P$ increases.  In particular, we find that in all cases 
$\varepsilon$ ranges from $\sim 10^{-1}$ for
$P=1$, to between $\sim 10^{-5}-10^{-7}$ for $P=11$, in all experiments.  As is evident from these tables, the error decreases uniformly with increase in $P$; a more detailed $P$ dependence can be gleaned from these Tables.  In Table \ref{off_p_convergence}, we demonstrate convergence in $P$ for the Helmholtz potential for 
$\mu=1$ and $\mu=2$ lattices in which sources are distributed outside of the dimension of periodicity, i.e. over a cube rather than a plane or a line.  In both cases, we find that the error convergence exhibits behavior similar to that found in 
Tables \ref{1d_p_convergence} and \ref{2d_p_convergence}.  One point of clarification is necessary in presenting convergence results for truncated infinite sums.  In some cases, the relative error in the potential apparently exceeds 
the relative error at which the sums are truncated ($\varepsilon_{GF}=10^{-5}$).  This is not anomalous in as far as the errors presented give an indication of how well the ACE expansions converge to the necessarily finite order approximation 
to the actual potential. Next, we present results for the Helmholtz kernel for $\mu=2$ subject to quasi-periodic boundary conditions, the details of which can be found in Appendix \ref{quasi_appendix}.  In 
Table \ref{quasi_error}, we find that at a fixed $P$, farfield error is largely independent of the quasi-periodic phase angle, in as far as error is observed to be of the same order of magnitude for all parameters.  This robustness 
opens our method to a number of applications in electromagnetics/optics, including oblique scattering from metamaterial structures or photonic crystal slabs. \newline

\begin{figure}
 \centering
 \includegraphics[scale=0.4]{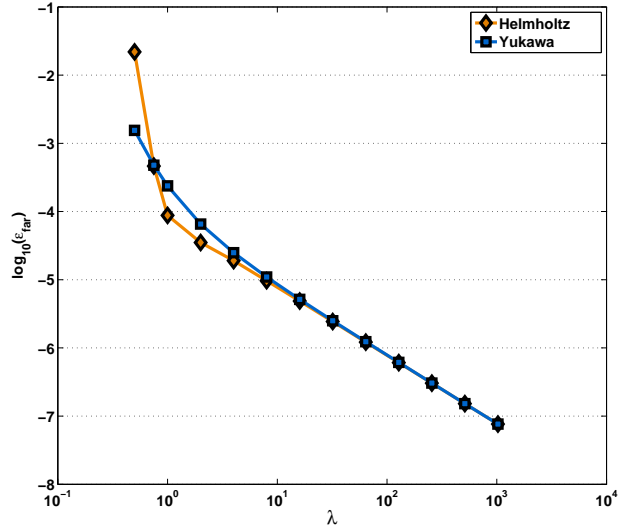}
 \caption{Error convergence as a function of $\lambda$ for 1000 points randomly distributed over a cube of unit volume for $\mu=2$ at a fixed value of $P=6$.} \label{k_figure}
\end{figure}

In demonstrating the accuracy and applicability of the ACE algorithm to a wide range of practical problems, it is interesting to analyze convergence of the ACE expansions at a fixed value of $P$ as $\kappa$ varies.
Recall that $\kappa$ is inversely proportional to the screening length for the Yukawa potential, and the wavelength for the Helmholtz potential, motivating the introduction of a unitless length scale, 
$\lambda = 2\pi/( \kappa (\uA_\mu)^{1/\mu}))$.  This scale roughly defines how rapidly the potential will vary over a unit cell, viz. the number of wavelengths in a unit cell for the Helmholtz potential.  That the ACE algorithm maintains a high degree of accuracy at a fixed value of $P$ 
for a broad range of relevant $\lambda$ values for both Helmholtz and Yukawa potentials is made evident in Fig. (\ref{k_figure}).  Here we have distributed 1000 unknowns over a cube of unit volume and evaluated 
$\varepsilon_{far}$, at a fixed value of $P=6$, for both the Helmholtz and Yukawa potentials with $\mu=2$ for $\lambda$ ranging from 0.5 to 1024. The error,
$\varepsilon_{far}$, decreases uniformly for $\lambda > 0.5$ from $\varepsilon_{far} \sim 10^{-1}-10^{-2}$ at $\lambda=0.5$ to $\varepsilon_{far} \sim 10^{-7}$ at $\lambda=1024$.  This is essentially due 
to the fact that as $\lambda$ increases, slowly varying terms become increasingly dominant in the Green's function.  That the error convergence appears to outperform the Coulombic ($\lambda \to \infty$) 
result in Table \ref{2d_p_convergence} is due to the fact that we have extracted the $\bn_\mu=0$ term from all Coulombic calculations, as discussed towards the end of 
Appendix \ref{gf_appendix}.  To examine the degradation in $\varepsilon_{far}$ as $\lambda$ decreases, at fixed $P$, it is useful to consider previous results and analysis for the non-periodic Helmholtz kernel, 
as presented in \cite{vikram09}.  Here, the authors allude to a decrease in the efficiency of the ACE algorithm at a fixed accuracy, i.e. as $\lambda$ becomes small relative to a box length, $P$ must be 
increased to achieve the same relative error.  This type of behavior is similarly evident in the periodic Helmholtz and Yukawa potentials.  In Table \ref{error_degradation}, $\varepsilon_{far}$ is calculated for a 
random distribution of 1000 points distributed over a planar unit cell ($\uA_\mu=1$) for both periodic and non-periodic \cite{vikram09} Helmholtz potentials at a fixed value of $P=8$.  This test is simply intended 
to demonstrate that the periodic and non-periodic ACE expansions degrade at approximately the same rate as wavelength is decreased at fixed $P$.  While increasing $P$ somewhat alleviates this, as $\lambda$ 
decreases, $P$ will become prohibitively large at some point in maintaining a desired level of accuracy, limiting the viability of ACE. \newline

To this end, it is worthwhile to examine this apparent limitation in the context of the physical relevance of small values of $\lambda$. Consider $\lambda = 0.5$ ($\kappa = 4\pi)$, the 
value for which $\epsilon_{far}$ is highest in Fig. (\ref{k_figure}).  For the Yukawa potential, $\lambda=0.5$ corresponds to a strongly screened system, in as far as second nearest neighbor cells 
will introduce a correction to the potential less than 0.001$\%$ of the correction due to the nearest neighbor cells.  In such a strongly screened system, the interactions are better handled by 
short-range acceleration techniques like linked-cell methods \cite{griebel09}.  Physically relevant values of $\lambda$ will instead be on the order of or greater than a primitive cell length ($\uA_\mu^{1/\mu}$), 
i.e. a weakly screened system for which the ACE algorithm is demonstrably accurate.  For the Helmholtz potential, rather than strong screening, small values of $\lambda$ correspond to 
increasingly oscillatory behavior.  Fortunately, most physically relevant conditions for the periodic Helmholtz potential involve values of $\lambda$ which correspond to systems in which the 
period of oscillation is on the order of, or greater than a primitive cell length.  Typically frequency selective structures, metamaterials, and photonic crystals, to name but a few applications, 
consist of subwavelength unit cells (i.e. $\lambda \geq 1$ on our scale), to which the ACE algorithm is also very well-suited.  

\subsection{Scaling}
As with error convergence, in demonstrating linear scaling we provide timings for the calculation of the farfield contributions to the total potential, and explicitly demonstrate that the remaining number of nearfield interactions 
will scale as $\mathcal{O}(N)$.  Tables \ref{1d_scaling}, \ref{2d_scaling}, and \ref{3d_scaling} contain a number of metrics which are indicative of the speed-up achieved 
in using the ACE algorithm.  These metrics are as follows: \newline
\begin{enumerate}
 \item {\bf $t_{ACE}$:} Time required for tree traversal, i.e. the execution of steps outlined in Algorithm \ref{tree_alg}
 \item {\bf $t_{pre}$:} Time required for the precomputation of ACE translation operators
 \item {\bf $N_{unique}$:} The number of unique translation operators that must be evaluated in precomputation
 \item {\bf $t_{far}$:} Time required for the direct evaluation of farfield interactions (i.e. computing and storing non-zero elements of $\bG_{\alpha\beta}^{far}=\bG_{\alpha\beta} - \bG_{\alpha\beta}^{near}$)
 \item {\bf $t_{direct}$:} Time required for the direct calculation of the farfield contribution to the potential (i.e. evaluation of $\bPsi^{direct}_\alpha$ as a matrix-vector product)
 \item {\bf $N_{near}$:} The remaining number of nearfield interactions (i.e. non-zero elements of $\bG_{\alpha\beta}^{near}$)
\end{enumerate}
\vspace{0.1in} In all 3 tables, potentials are evaluated for $P=7$, yielding a farfield error $\mathcal{O}(10^{-5})$.  Timings with prepended tildes were extrapolated 
based upon multiplication by a factor determined by the increase in the number of farfield interactions.  In all problems, the primitive cells are of unit 
dimension, and for the Helmholtz and Yukawa potentials, physically realistic values of $\kappa$ ($2\pi$ and $2\pi/10$, respectively) were chosen.  All Ewald sums were 
evaluated to a relative error of $10^{-5}$, in both ACE and direct calculations. In all tests, the average number of unknowns per leaf box is fixed at 64. \newline       

In all cases, a considerable speedup in evaluating the farfield contribution to the potential is achieved, ranging from a factor of 12 for 1024 unknowns with $\mu=1$ to 
6 million for $\sim 1$ million unknowns with $\mu=2$.  In all cases, the breakeven point is found to be rather low; 540 unknowns for $\mu=1$, 570 unknowns for $\mu=2$, and 
1730 unknowns for $\mu=3$.  It is evident in all 3 tables that $\mathcal{O}(N)$ scaling is achieved in tree traversal.  A linear regression on a log-log scale yields a scaling exponent 
which differs from 1 by less than 3\% in all cases under consideration.  Similarly, linear scaling is apparent for $N_{near}$, with a scaling exponent that is also very nearly unity; 
the specific exponent is given in the caption to the Tables.  Finally, the precomputation time, $t_{pre}$ exhibits clear sublinear scaling in $N$ for all cases.  The cost of the precomputation is typically ignored in the literature, usually being 
dismissed as a one time cost, negligible relative to the repeated calculation of potentials from the same tree.  While this is often the case for non-periodic potentials, 
as the evaluation of the requisite periodic translation operators is seemingly non-trivial (i.e. many infinite sums), it is important to demonstrate that this step can be completed 
on a time scale that is not prohibitively long, relative to a single tree traversal.  One final point worth noting is that we have intentionally chosen a non-optimal value 
for the density of unknowns per leaf box at $P=7$.  Cost is optimized for a density of $\sim 20$ unknowns per box, whereas our tests were run at 64 unknowns per box.  As there is less 
control over the density of points per leaf box, due to the manner in which the tree is constructed (i.e. number of levels instead of leaf box size), it is important to demonstrate 
that both scaling and a low breakeven point are maintained for densities away from the optimum.  At more optimal densities, the scaling exponent will be closer to $1$ and 
the breakeven points will decrease, as well.

\section{Conclusion}
In this paper, we have presented extensions of the ACE algorithm which realize the $\mathcal{O}(N)$ calculation of Helmholtz, Yukawa, and Coulomb potentials on singly, 
doubly, and triply periodic lattices.  The results presented demonstrate error convergence as well as considerable acceleration.  Further work is being submitted elsewhere to 
demonstrate the applicability of our method to the analysis of electromagnetic wave propagation \cite{baczewski11}.  We are presently working on adapting these techniques to 
solid-state electronic structure calculations involving defects.  

\section{Acknowledgements}
The authors would like to thank Melapudi Vikram, Naveen Nair, and He Huang for countless useful discussions.  A.B. would like to thank the National Science Foundation Graduate Research Fellowship for funding his 
graduate studies.  More broadly, this work has been funded by NSF CCF-0729157 and NSF DMS-0811197.  Finally, we thank the High Performance Computing Center (HPCC) at Michigan State University for access to computational 
resources.

\begin{appendix}

\section{Representations of the Periodic Green's Function}\label{gf_appendix}

In this Appendix, we discuss different representations of $G_{\mu}(|\br-\br'|)$, in particular the direct, spectral, and Ewald forms.  
Perhaps the simplest of these is the direct representation, in which $G_{\mu}(|\br-\br'|)$ is furnished by a sum of the relevant free 
space Green's function, $G(|\br-\br'|)$ over $\Lmu$:
\begin{equation}
 \label{direct_gf}
 G_{\mu}(|\br-\br'|) = \displaystyle\sum_{\bt(\bn_\mu) \in \Lmu} G(|\br-\br'+\bt(\bn_\mu)|)
\end{equation}
For the PDEs specified in (\ref{pdes}), the proper free space Green's functions are:
\begin{subequations}
\label{gfs}
\begin{align}
&G(|\br-\br'|) = \frac{e^{-i\kappa|\br-\br'|}}{4\pi |\br-\br'|}~~~\text{(Helmholtz~Equation)} \\
&G(|\br-\br'|) = \frac{e^{-\kappa|\br-\br'|}}{4\pi |\br-\br'|}~~~\text{(Yukawa~Equation)} \\
&G(|\br-\br'|) = \frac{1}{4\pi |\br-\br'|}~~~\text{(Poisson~Equation)}
\end{align}
\end{subequations}
It is evident that, for the Helmholtz  (\ref{direct_gf}) is conditionally convergent, and for the Coulomb potential it is manifestly divergent (a topic which is discussed in more detail, later). 
While this is not the case for the Yukawa potential, for very small values of $\kappa$ (i.e. $1/\kappa >>$ max$\lbrace |\lv_i|~| i \in 1,..,\mu \rbrace$), (\ref{direct_gf}) may be slowly convergent relative to other 
representations.  Alternatively, we might pursue a spectral representation of $G_\mu(|\br-\br'|)$, in which case we transform the sum over $\Lmu$, to a summation over its reciprocal lattice, $\Lmu^*$.  This is typically 
accomplished by way of the Poisson summation formula:

\begin{equation}
 \label{poisson_sum}
 \displaystyle\sum_{\bt(\bn_\mu)} f(\br+\bt(\bn_\mu)) = \frac{1}{A_\mu} \displaystyle\sum_{\bk(\bn_\mu)} e^{i\bk(\bn_\mu)\cdot \br} \mathcal{F}\left(\bk(\bn_\mu)\right),~\text{where}~\mathcal{F}\left(\bk(\bn_\mu)\right)=\int d^\mu\br' e^{-i\bk(\bn_\mu)\cdot\br'} f(\br')
\end{equation}

Typically, this identity is applied to Eqn. (\ref{direct_gf}), directly yielding the spectral representation of the periodic Green's function.  Rigorously, however, this is only permissible for the Yukawa potential, as the 
direct representation is a conditionally convergent sum.  Uniform convergence of the summation on the left hand side of (\ref{poisson_sum}) is in fact a necessary condition for this identity's validity \cite{walker88}.  
This deficiency can be circumvented by inserting a convergence factor, as in \cite{deleeuw80}, which yields effectively the same result as the Poisson summation, excepting singular contributions which need be regularized 
depending upon the application.

\begin{subequations}
\label{spectral_gf}
\begin{align}
 &G_1(|\br-\br'|) = \frac{1}{2\pi \uA_1}\displaystyle\sum_{\bk(\bn_1) \in \Lma^*} e^{i\bk(\bn_1) \cdot \br_l} K_0(\alpha(\bn_1) |\br_t|)~~~&\text{($\mu=1$)}\\
 &G_2(|\br-\br'|) = \frac{1}{2 \uA_2}\displaystyle\sum_{\bk(\bn_2) \in \Lmb^*} e^{i\bk(\bn_2) \cdot \br_l} \frac{e^{-\alpha(\bn_2)|\br_t|}}{\alpha(\bn_2)}~~~&\text{($\mu=2$)} \\
 &G_3(|\br-\br'|) = \frac{1}{\uA_3}\displaystyle\sum_{\bk(\bn_3) \in \Lmc^*} e^{i\bk(\bn_3) \cdot \br_l} \frac{1}{\alpha(\bn_3)^2}~~~&\text{($\mu=3$)}
\end{align}
\end{subequations}
Here, $\br_l$ and $\br_t$ are the components of $\br-\br'$ which lie in and out of the lattice, respectively, and the functional form of $\alpha(\bn_\mu)$ is given in Eqn. (\ref{spectral_param}).
For $\mu = 1$ or $2$, the sums in (\ref{spectral_gf}) exhibit spectral convergence away from the lattice, i.e. $|\br_t| > 0$.  However, for $|\br_t| \to 0$ or $\mu=3$, these sums have poor 
convergence properties which considerably limits their numerical utility. \\

To achieve a representation of $G_\mu(|\br-\br'|)$ which is absolutely and rapidly convergent both near and far from the lattice, we turn to Ewald's method \cite{ewald21}.  In the Ewald representation, 
the Green's function is separated into two rapidly convergent sums, one on $\Lmu$, the other on $\Lmu^*$.  Our derivation begins with the following integral representations of the free space Green's functions:
\begin{subequations}
\label{integral_identity}
 \begin{align}
  &G(|\br-\br'|) = \frac{1}{2\pi^{3/2}} \int \limits_0^\infty ds~e^{-s^2|\br-\br'|^2+\frac{\kappa^2}{4s^2}}~&\text{(Helmholtz)} \\
  &G(|\br-\br'|) = \frac{1}{2\pi^{3/2}} \int \limits_0^\infty ds~e^{-s^2|\br-\br'|^2-\frac{\kappa^2}{4s^2}}~&\text{(Yukawa)} \\
  &G(|\br-\br'|) = \frac{1}{2\pi^{3/2}} \int \limits_0^\infty ds~e^{-s^2|\br-\br'|^2}~&\text{(Poisson)}
\end{align}
\end{subequations}
One means of arriving at the Ewald representation is to split this integration on $\left[0,\infty \right)$ into separate integrals on $\left[0,\eta \right]$ and $\left[\eta,\infty \right)$, where $\eta \in \mathbb{R}^{+}$.
We can then represent the periodic Green's function as:
\begin{equation} 
 \label{ewald1}
 G_{\mu}(|\br-\br'|) = \sum \limits_{\bt(\bn_\mu) \in \Lmu}~\int \limits_0^\eta ds~\mathcal{G}(s,|\br-\br'+\bt(\bn_\mu)|) + \sum \limits_{\bt(\bn_\mu) \in \Lmu}~\int  \limits_\eta^\infty ds~\mathcal{G}(s,|\br-\br'+\bt(\bn_\mu)|)
\end{equation}
Here, the form of $\mathcal{G}(s,|\br-\br'+\bt(\bn_\mu)|)$ is evident from Eqn. (\ref{integral_identity}).  We exchange summation and integration, and then apply the identity in 
Eqn. (\ref{poisson_sum}) to the first integral:
\begin{equation}
 \sum \limits_{\bt(\bn_\mu) \in \Lmu}~\int \limits_0^\eta ds~\mathcal{G}(s,|\br-\br'+\bt(\bn_\mu)|) = \sum \limits_{\bk(\bn_\mu) \in \Lmu^*}~\int \limits_0^\eta ds~\int d^\mu\br'' e^{-i\bk(\bn_\mu)\cdot\br''}\mathcal{G}(s,|\br''|)
\end{equation}
Evaluating all integrals, we are left with the Ewald representation of the periodic Green's function:
\begin{equation}
  G_{\mu}(|\br-\br'|) = \sum \limits_{\bk(\bn_\mu) \in \Lmu^*}\Ewk(\br-\br',\bk(\bn_\mu)) + \sum \limits_{\bt(\bn_\mu) \in \Lmu}\Ewr(|\br-\br'+\bt(\bn_\mu)|) 
\end{equation}
The functional forms of $\Ewr(|\br-\br'+\bt(\bn_\mu)|)$ and $\Ewk(\br-\br',\bk(\bn_\mu))$ are given in Eqns. (\ref{real_ewald}) and (\ref{recip_ewald}). \\

While we ignore singular contributions to the potential throughout this paper, we provide a brief discussion of them here for the sake of completeness.  The most evident singular contributions come about due 
to the so-called self-terms (i.e. $|\br-\br'| \to 0$) which we have subtracted out in Eqn. (\ref{mat-vec}).  The exact manner in which this is regularized in practice is largely application dependent.  In calculating
potential energies in a Coulombic system, a Laurent expansion of the Green's function is employed, and the term which scales as $\frac{1}{|\br-\br'|}$ is simply negated \cite{deleeuw80}.  In the context of some integral 
equation discretization schemes, such as the Method of Moments, the singularity is regularized by the integration measure associated with the source and testing integrals. \\

Another type of singular contribution can arise due to the situation in which $\alpha(\bn_\mu) \to 0$.  Again, the manner in which these behaviors are regularized are application dependent.  In Coulombic systems, 
this singular contribution vanishes for charge neutral primitive cells, up to a correction proportional to the dipole moment of the primitive cell \cite{deleeuw80}.  As one of our applications of interest is in 
studying electronic defects in which the primitive cell is not charge neutral, we have not guaranteed charge neutrality in numerical experiments involving Coulombic potentials.  Instead, we simply ignore the 
$\alpha(\bn_\mu) \to 0$ term of the Ewald representation of the periodic Coulombic Green's function.  As this contribution is spatially uniform, it does not affect our error convergence.

\section{Evaluation of Ewald Sums}\label{gf_calc}

In evaluating the periodic Green's function in the Ewald representation, as well as the periodic ACE translation operators which have the form of an Ewald sum, a few 
considerations are necessary to achieve optimal results, namely (i) the order in which terms are added, (ii) criteria for truncating the summation, and (iii) the choice 
of an appropriate splitting parameter, $\eta$.  The calculation of $G_{\mu}(|\br-\br'|)$ begins with the contribution due to $\bn_\mu = 0$, and subsequent terms are 
added on over surfaces of constant $|\bn_\mu|$ (points $\to \mu=1$, circles $\to \mu=2$, and spherical shells $\to \mu=3$) for increasing $|\bn_\mu|$.  We denote the 
partial sum over the surface for 
which $|\bn_{\mu}|=m$ as: 
\begin{equation}
  G_\mu(|\br-\br'|)|_m = \displaystyle\sum_{\substack{\bt(\bn_\mu) \\ |\bn_\mu|=m}} \Ewr(|\br-\br'+\bt(\bn_\mu)|) + 
                         \displaystyle\sum_{\substack{\bk(\bn_\mu) \\ |\bn_\mu|=m}} \Ewk(\br-\br',\bk(\bn_\mu))
\end{equation}
Such that:
\begin{equation}
 G_\mu(|\br-\br'|) = \sum \limits_{m=0}^{\infty} G_\mu(|\br-\br'|)|_m
\end{equation}
Our criterion for the convergence of this sum is given in terms of the relative convergence of the $M$th partial sum in the $L_2$ norm: 
\begin{equation}
 \sqrt{\frac{\left|\sum \limits_{m=0}^{M} G_\mu(|\br-\br'|)|_m - \sum \limits_{m=0}^{M-1}G_\mu(|\br-\br'|)|_{m}\right|^2}{\left|\sum \limits_{m=0}^{M} G_\mu(|\br-\br'|)|_m\right|^2}} \leq \varepsilon_{GF}
\end{equation}
In all of the results presented in this paper, $\varepsilon_{GF} = 10^{-5}$ for both the direct evaluation of the Ewald sum, as well as the ACE translation operators. \\
  
The only remaining consideration is $\eta$, which controls the relative rate of convergence of the real and reciprocal sums.  As $\eta$ increases, the contribution of the 
reciprocal sum to the overall convergence of the Ewald sum is increased.  In the literature, an optimal value of $\eta$ is considered to be the one for which the 
reciprocal and real sums have the same asymptotic rate of convergence \cite{capolino07,jordan86,kustepeli00}.  This optimal value of $\eta$ has different forms for different values of $\mu$.  
\begin{subequations}
\begin{align}
 \eta_{opt} & = \sqrt{\pi} |\lv_1|^{-1}~~~&(\mu=1) \\
	    & = \sqrt{\pi} (|\lv_1||\lv_2|)^{-1/2}~~~&(\mu=2) \\
            & = \sqrt{\pi} \left(\frac{|\lv_1|^{-2} + |\lv_2|^{-2} + |\lv_3|^{-2}}{|\lv_1|^2 + |\lv_2|^2 + |\lv_3|^2}\right)^{1/4}~~~&(\mu=3) \\
\end{align}
\end{subequations}
In practice, we have found that using the same $\eta$ for the Green's function itself and the ACE translation operators delivers ideal performance in terms of both 
speed and accuracy.  Unless otherwise indicated, we employ this optimal value for $\eta$ in all numerical experiments.  There are conditions 
under which alternative values of $\eta$ must be used for the Helmholtz potential.  It is well-established in the literature that using $\eta_{opt}$ for the periodic Helmholtz potential will lead to 
a `catastrophic loss' in accuracy, due to finite precision arithmetic, for situations in which the unit cell is on the order of, or larger than a wavelength.  Methods to 
mitigate this loss in accuracy at the expense of sub-optimal convergence have been proposed \cite{capolino07,lovat08,oroskar06}, and are utilized when appropriate.

\section{Derivation of Real Sum ACE Translation Operators}\label{trans_ops}

In this Appendix, we derive the expressions given in Eqn. (\ref{real_sum_expressions}) for the translation operator components arising due to the real sum in the Ewald representation of $G_\mu(|\br-\br'|)$:
\begin{equation}
 \nabla^{(p)}\Ewr(|\br_o^p-\br_s^p+\bt(\bn_\mu)|)\left[p_x,p_y,p_z \right] = \partial_x^{p_x} \partial_y^{p_y} \partial_z^{p_z}\Ewr(|\br_o^p-\br_s^p+\bt(\bn_\mu)|)
\end{equation}
While straightforward partial differentiation of the expression given in Eqn. (\ref{real_ewald}) will yield a viable expression, we find that a more computationally efficient expression can be arrived 
at by manipulating the integral representation given in Eqn. (\ref{ewald1}).  For the periodic Helmholtz potential, we proceeed as follows:

\begin{subequations}
\begin{align}
  \nabla^{(p)} \Ewr(\bR)\left[p_x, p_y, p_z \right] &= \frac{1}{2\pi^{3/2}} \int\limits_\eta^\infty ds~\partial_x^{p_x}\partial_y^{p_y}\partial_z^{p_z}e^{-s^2|\bR|^2+\frac{\kappa^2}{4s^2}} \\
                                \label{h_prod}&= \frac{1}{2\pi^{3/2}} \int\limits_\eta^\infty ds~(-s)^p H_{px}(sR_x)H_{py}(sR_y)H_{pz}(sR_z)e^{-s^2|\bR|^2+\frac{\kappa^2}{4s^2}} \\
					      &= \frac{(-1)^p}{2\pi^{3/2}} \int\limits_\eta^\infty ds~\displaystyle\sum_{m=0}^{p} C_{m}^{p_x,p_y,p_z} s^{p+m} e^{-s^2|\bR|^2+\frac{\kappa^2}{4s^2}} 
\end{align}
\end{subequations}
Here, $C_m^{p_x,p_y,p_z}$ is the coefficient of the term which is $m$th order in $s$ in the product of Hermite polynomials in Eqn. (\ref{h_prod}).  To evaluate this integral in closed form, we expand 
the $\exp(\kappa^2/4s^2)$ term in a Laurent series in $s$.
\begin{subequations}
\begin{align}
  \nabla^{(p)} \Ewr(|\bR|)\left[p_x, p_y, p_z \right] &= \frac{(-1)^p}{2\pi^{3/2}} \int\limits_\eta^\infty ds~\displaystyle\sum_{m=0}^{p} C_{m}^{p_x,p_y,p_z} \sum_{\mu=0}^{\infty} \frac{(\kappa/2s)^{2\mu}}{\mu!} s^{p+m} e^{-s^2|\bR|^2} \\
					      &= \frac{(-1)^p}{2\pi^{3/2}} \displaystyle\sum_{m=0}^{p} \sum_{\mu=0}^{\infty} C_{m}^{p_x,p_y,p_z} \int\limits_\eta^\infty ds~ \frac{(\kappa/2s)^{2\mu}}{\mu!} s^{p+m} e^{-s^2|\bR|^2} \\
					      &= \frac{(-1)^p}{4\pi^{3/2}|\bR|} \displaystyle\sum_{m=0}^{p} \sum_{\mu=0}^{\infty} C_{m}^{p_x,p_y,p_z} \frac{(\kappa^2 |\bR|^2/4)^{\mu}}{\mu!} \frac{\Gamma\left(\frac{p+m+1}{2}-\mu, \eta^2|\bR|^2 \right)}{|\bR|^{p+m}}
\end{align}
\end{subequations}
Here, $\Gamma(n,x)$ is the incomplete Gamma function of $n$th order.  These steps can be repeated for the Yukawa and Coulomb potentials, yielding the following expressions:
\begin{subequations}
 \begin{align}
  \nabla^{(p)} \Ewr(|\bR|)\left[p_x, p_y, p_z \right] &= \frac{(-1)^p}{4\pi^{3/2}|\bR|} \displaystyle\sum_{m=0}^{p} \sum_{\mu=0}^{\infty} C_{m}^{p_x,p_y,p_z} \frac{(\kappa^2 |\bR|^2/4)^{\mu}}{\mu!} \frac{\Gamma\left(\frac{p+m+1}{2}-\mu, \eta^2|\bR|^2 \right)}{|\bR|^{p+m}} &\text{(Helmholtz)}\\
					      &= \frac{(-1)^p}{4\pi^{3/2}|\bR|} \displaystyle\sum_{m=0}^{p} \sum_{\mu=0}^{\infty} C_{m}^{p_x,p_y,p_z} \frac{(-\kappa^2 |\bR|^2/4)^{\mu}}{\mu!} \frac{\Gamma\left(\frac{p+m+1}{2}-\mu, \eta^2|\bR|^2 \right)}{|\bR|^{p+m}} &\text{(Yukawa)} \\ 
					      &= \frac{(-1)^p}{4\pi^{3/2}|\bR|} \displaystyle\sum_{m=0}^{p} C_{m}^{p_x,p_y,p_z} \frac{\Gamma\left(\frac{p+m+1}{2}, \eta^2|\bR|^2 \right)}{|\bR|^{p+m}} &\text{(Coulomb)}
 \end{align}
\end{subequations}
All sums over incomplete Gamma functions are rapidly convergent for physically relevant values of the arguments, and recurrence relations are utilized to rapidly compute each term.  

\section{Derivation of Error Bounds on ACE Expansions}\label{error_appendix}

\subsection{Bounds on Terms in the Real Sum for the Coulomb Potential}
Using Eqns. (\ref{coulomb_real_ace}) and (\ref{m2l_error}), the error incurred in approximating the $\bn_\mu$th term of the real sum by truncating the M2L expansion 
above $P$th order is given as:
\begin{equation}
  \label{bound_start}
  \varepsilon_{m,r}(\bn_\mu) \leq \left|\displaystyle\sum_{n=P+1}^{\infty} \bM^{(n)} \cdot n \cdot \frac{(-1)^n}{4\pi^{3/2}|\bR|} \displaystyle\sum_{m=0}^{n} \bC_{m}^{(n)} \frac{\Gamma\left(\frac{n+m+1}{2}, \eta^2|\bR|^2 \right)}{|\bR|^{n+m}}\right|
\end{equation}
Here, $\bC^{(n)}_{m}$ is a tensor whose components consist of the products of Hermite polynomials given in Eqn. (\ref{h_prod}).  For an arbitrary $n$th rank tensor $\bA^{(n)}$, the following inequality holds:
\begin{equation}
 \left|\bA^{(n)} \cdot n \cdot \bC^{(n)}_m\right| \leq C_H \left|\bA^{(n)} \cdot n \cdot \frac{\bR^{(n)}}{|\bR|^{n-m}}\right|
\end{equation}
As $\Gamma(n,x)$ increases monotonically in $n$, the following simple inequality will hold:
\begin{equation}
 \sum_{m=0}^{n} \bR^{(n)} \frac{\Gamma\left(\frac{n+m+1}{2}, \eta^2|\bR|^2 \right)}{|\bR|^{2n}} \leq C \bR^{(n)}~n~\frac{\Gamma\left(n+\frac{1}{2}, \eta^2|\bR|^2 \right)}{|\bR|^{2n}}
\end{equation}
Combining the previous two inequalities, we can manipulate Eqn. (\ref{bound_start}) into the following form:
\begin{equation}
  \label{bound_mid}
  \varepsilon_{m,r}(\bn_\mu) \leq C \left|\displaystyle\sum_{n=P+1}^{\infty} \bM^{(n)} \cdot n \cdot \bR^{(n)} \frac{(-1)^n n \Gamma\left(n+\frac{1}{2}, \eta^2|\bR|^2 \right)}{4\pi^{3/2}|\bR|^{2n+1}} \right|
\end{equation}
Next, we recall the following inequality for an arbitrary $n$th rank tensor, $\bA^{(n)}$, contracted with a Multipole expansion in which the furthest point source from the 
origin is at position $\br_{i,max}$ \cite{shanker07}:
\begin{equation}
\label{rmax_ineq}
|\bA^{(n)} \cdot n \cdot \bM^{(n)}| \leq C \frac{1}{n!} |\bA^{(n)} \cdot n \cdot \br_{i,max}^{(n)}|
\end{equation}
Applying this inequality to Eqn. (\ref{bound_mid}):
\begin{equation}
  \varepsilon_{m,r}(\bn_\mu) \leq C \left|\displaystyle\sum_{n=P+1}^{\infty} \frac{\br_{i,max}^{(n)}}{n!} \cdot n \cdot \bR^{(n)} \frac{(-1)^n n \Gamma\left(n+\frac{1}{2}, \eta^2|\bR|^2 \right)}{4\pi^{3/2}|\bR|^{2n+1}} \right|
\end{equation}
Using the integral representation of the incomplete Gamma function and exchanging summation and integration, we are left with the following:
\begin{equation}
  \varepsilon_{m,r}(\bn_\mu) \leq C \left|\displaystyle \int \limits_{\eta^2|\bR|^2}^{\infty} dt \sum_{n=P+1}^{\infty} \frac{\br_{i,max}^{(n)}}{n!} \cdot n \cdot \bR^{(n)} \frac{(-1)^n n t^{n-1/2} e^{-t}}{4\pi^{3/2}|\bR|^{2n+1}} \right|
\end{equation}
Applying what is essentially the Cauchy-Schwartz inequality to $\br_{i,max}^{(n)}\cdot n \cdot \bR^{(n)}$, as in \cite{shanker07}:
\begin{equation}
  \varepsilon_{m,r}(\bn_\mu) \leq C \left| \frac{(|\br_{i,max}||\bR|)^{P+1}}{4\pi^{3/2} (P+1)! |\bR|^{2P+3}} \displaystyle \int \limits_{\eta^2|\bR|^2}^{\infty} dt~t^{P+1/2} e^{-t} \sum_{n=0}^{\infty} \frac{(n+P+1)}{n!} \frac{(-1)^n (|\br_{i,max}||\bR|)^{n} t^{n}}{|\bR|^{2n}} \right|
\end{equation}
Resolving the infinite sum inside of the integrand:
\begin{equation}
  \varepsilon_{m,r}(\bn_\mu) \leq C \left| \frac{|\br_{i,max}|^{P+1}}{4\pi^{3/2} (P+1)! |\bR|^{P+2}} \displaystyle \int \limits_{\eta^2|\bR|^2}^{\infty} dt~t^{P+1/2} \left(P+1-\frac{|\br_{i,max}|}{|\bR|}t\right)e^{-\left(1+\frac{|\br_{i,max}|}{|\bR|}\right)t} \right|
\end{equation}
We adopt the notation, $a=\frac{|\br_{i,max}|}{|\bR|}$, and note that $a \leq \sqrt{3}/4$.  In the ACE algorithm, $|\bR| \geq 2dx_0$, based upon the criterion for `farfield' interactions and 
$|\br_{i,max}| \leq \sqrt{3}/2 dx_0$, for a box size of $dx_0$.  Evaluating this integral, we arrive at the following final expression for our bound:
\begin{equation}
  \varepsilon_{m,r}(\bn_\mu) \leq C \frac{a^{P+1}}{4\pi^{3/2} (P+1)! |\bR|} \left|(P+1)\frac{\Gamma(P+3/2,(1+a)\eta^2|\bR|^2)}{(1+a)^{P+3/2}} - a\frac{\Gamma(P+5/2,(1+a)\eta^2|\bR|^2)}{(1+a)^{P+5/2}} \right|
\end{equation}
We note that a looser, but monotonically decreasing bound is given by:
\begin{equation}
  \varepsilon_{m,r}(\bn_\mu) \leq C \frac{a^{P+1}}{4\pi^{3/2} (P+1)! |\bR|} \left|(P+1)\frac{\Gamma(P+3/2)}{(1+a)^{P+3/2}} - a\frac{\Gamma(P+5/2)}{(1+a)^{P+5/2}} \right|
\end{equation}

\subsection{Bounds on Terms in the Reciprocal Sum for $\mu=3$}

Starting from the expression for $\nabla^{P}\Ewk(|\bR|)$ given in Eqn. (\ref{top_k_3}):
\begin{subequations}
 \begin{align}
  \varepsilon_{m,k}(\bn_3) &\leq \left|\displaystyle\sum_{n=P+1}^{\infty} \bM^{(n)} \cdot n \cdot (i\bk(\bn_3))^{(n)} \frac{e^{i\bk(\bn_3)\cdot\bR_l-\alpha^2(\bn_3)/4\eta^2}}{\uA_3 \alpha^2(\bn_3)} \right| \\
               &\leq C \left|\frac{e^{-\alpha^2(\bn_3)/4\eta^2}}{\uA_3 \alpha^2(\bn_3)} \right| \left|\displaystyle\sum_{n=P+1}^{\infty} \bM^{(n)} \cdot n \cdot (i\bk(\bn_3))^{(n)} \right| 
 \end{align}
\end{subequations}
 Using the inequality in Eqn. (\ref{rmax_ineq}):
\begin{equation}
   \varepsilon_{m,k}(\bn_3) \leq C \frac{(\bk(\bn_3)\cdot\br_{i,max})^{P+1}}{(P+1)!} \left|\frac{e^{-\alpha^2(\bn_3)/4\eta^2}}{\uA_3 \alpha^2(\bn_3)} \right|
\end{equation}

\section{Quasi-Periodic Boundary Conditions}\label{quasi_appendix}

Quasi-periodic boundary conditions typically arise when considering Helmholtz-type problems in which an array of scatterers is excited at oblique incidence.  
We characterize this excitation in terms of a plane wave of the form $exp(i\bk_0\cdot\br)$, where $|\bk_0|=\kappa$, the effect of which is manifest as a phase factor 
applied to the potential when its argument is translated by a lattice vector.
\begin{equation}
 \label{qbc_psi}
 \psi(\br + \bt(\bn_\mu)) = e^{i\bk_0\cdot\bt(\bn_\mu)}\psi(\br):\bt(\bn_\mu) \in \Lmu
\end{equation}
This stands in contrast to the standard periodic boundary condition given in Eqn. (\ref{pbc_psi}).  This modification of the boundary conditions simply necessitates 
the use of a quasi-periodic Green's function and its associated translation operator.  No further modifications of the presented algorithm are necessary to give consideration 
to this class of potentials.  The quasi-periodic Green's function, $G_{\mu,\bk_0}(|\br-\br'|)$, can be written in terms of the Ewald representation of the periodic Green's 
function (Eqns. (\ref{real_ewald}) and (\ref{recip_ewald})) as follows:
\begin{align}
  G_{\mu,\bk_0}(|\br-\br'|) = \displaystyle\sum_{\bt(\bn_\mu)} e^{i\bk_0\cdot\bt(\bn_\mu)}\Ewr(|\br-\br'+\bt(\bn_\mu)|) + 
                      \displaystyle\sum_{\bk(\bn_\mu)} \Ewk(\br-\br',\bk(\bn_\mu)+\bk_0)
\end{align}
This same transformation can be applied to the expressions for the periodic translation operator (Eqns. (\ref{real_sum_expressions}) and (\ref{rec_sum_expressions})) 
to arrive at expressions for the quasi-periodic translation operator.  We note that while we only present results for the quasi-periodic Helmholtz potential, this 
approach is sufficiently general that it can be applied to quasi-periodic Coulomb or Yukawa potentials, should these expressions be relevant for some application.

\end{appendix}

\bibliographystyle{elsart-num}
\bibliography{periodic_bib}

\begin{thebibliography}{10}
\expandafter\ifx\csname url\endcsname\relax
  \def\url#1{\texttt{#1}}\fi
\expandafter\ifx\csname urlprefix\endcsname\relax\def\urlprefix{URL }\fi

\bibitem{otani08a}
Y.~Otani, N.~Nishimura, A periodic fmm for maxwell's equations in 3d and its
  application to problems related to photonic crystals, Journal of
  Computational Physics 227 (2008) 4630--4652.

\bibitem{otani09}
Y.~Otani, N.~Nishimura, An fmm for orthotropic periodic boundary value problems
  for maxwell's equations, Waves in Random and Complex Media 19 (2009) 80--104.

\bibitem{munk00}
B.~Munk, Frequency Selective Surfaces: Theory and Design, Wiley-Interscience,
  2000.

\bibitem{springel05a}
V.~Springel, The cosmological simulation code gadget-2, Mon. Not. R. Astron.
  Soc. 364 (2005) 1105--1134.

\bibitem{hung09}
L.~Hung, E.~Carter, Accurate simulations of metals at the mesoscale: Explicit
  treatment of 1 million atoms with quantum mechanics, Chemical Physics Letters
  475 (2009) 163--170.

\bibitem{shin09}
I.~Shin, A.~Ramasubramaniam, C.~Huang, L.~Hung, E.~Carter, Orbital-free density
  functional theory simulations of dislocations in aluminum, Philosophical
  Magazine 89 (2009) 3195--3213.

\bibitem{barnes86}
J.~Barnes, P.~Hut, A hierarchical $\mathcal{O}(n\log(n))$ force-calculation
  algorithm, Nature 324 (1986) 446--449.

\bibitem{duan00}
Z.~Duan, R.~Krasny, An ewald summation based multipole method, Journal of
  Chemical Physics 113 (2000) 3492--3495.

\bibitem{greengard87}
L.~Greengard, V.~Rokhlin, A fast algorithm for particle simulations, Journal of
  Computational Physics 73 (1987) 325--348.

\bibitem{schmidt91}
K.~Schmidt, M.~Lee, Implementing the fast multipole method in three dimensions,
  Journal of Statistical Physics 63 (1991) 1223--1235.

\bibitem{hockney_book}
R.~Hockney, J.~Eastwood, Computer Simulation Using Particles, Taylor \&
  Francis, 1989.

\bibitem{izaguirre05}
J.~Izaguirre, S.~Hampton, T.~Matthey, Parallel multigrid summation for the
  n-body problem, Journal of Parallel and Distributed Computing 65 (2005)
  949--962.

\bibitem{hockney73}
R.~Hockney, S.~Goel, J.~Eastwood, A 10000 particle molecular dynamics model
  with long range forces, Chemical Physics Letters 21 (1973) 589--591.

\bibitem{griebel09}
M.~Griebel, S.~Knapek, G.~Zumbusch, Numerical Simulation in Molecular Dynamics,
  Springer-Verlag, 2007.

\bibitem{cheng99}
H.~Cheng, L.~Greengard, V.~Rokhlin, A fast adaptive multipole algorithm in
  three dimensions, Journal of Computational Physics 155 (1999) 468--498.

\bibitem{li09}
P.~Li, H.~Johnston, R.~Krasny, A cartesian treecode for screened coulomb
  interactions, Journal of Computational Physics 228 (2009) 3858--3868.

\bibitem{shanker07}
B.~Shanker, H.~Huang, Accelerated cartesian expansions - a fast method for
  computing of potentials of the form r$^{-\nu}$ for all real $\nu$, Journal of
  Computational Physics 226 (2007) 732--753.

\bibitem{vikram07}
M.~Vikram, B.~Shanker, Fast evaluation of time domain fields in sub-wavelength
  source/observer distributions using accelerated cartesian expansions (ace),
  Journal of Computational Physics 227 (2007) 1007--1023.

\bibitem{vikram10}
M.~Vikram, A.~Baczewski, B.~Shanker, L.~Kempel, Accelerated cartesian expansion
  (ace) based framework for the rapid evaluation of diffusion, lossy wave, and
  klein-gordon potentials, Journal of Computational Physics UPDATE (2010) .

\bibitem{vikram09}
M.~Vikram, H.~Huang, B.~Shanker, T.~Van, A novel wideband fmm for fast integral
  equation solution of multiscale problems in electromagnetics, IEEE
  Transactions on Antennas and Propagation 57 (2009) 2094--2104.

\bibitem{nishimura02}
N.~Nishimura, Fast multipole accelerated boundary integral equation methods,
  Applied Mechanics Reviews 55 (2002) 299--324.

\bibitem{vikram_review09}
M.~Vikram, B.~Shanker, An incomplete review of fast multipole methods - from
  static to wideband - as applied to problems in computational
  electromagnetics, ACES 24.

\bibitem{schmidt97}
K.~Schmidt, M.~Lee, Multilevel ewald sums for the fast multipole method,
  Journal of Statistical Physics 89 (1997) 411--424.

\bibitem{challacombe97}
M.~Challacombe, C.~White, M.~Head-Gordon, Periodic boundary conditions and the
  fast multipole method, Journal of Chemical Physics 107 (1997) 10131--10140.

\bibitem{lambert96}
C.~Lambert, T.~Darden, J.~Board, A multipole-based algorithm for efficient
  calculation of forces and potentials in macroscopic periodic assemblies of
  particles, Journal of Computational Physics 126 (1996) 274--285.

\bibitem{rokhlin94}
V.~Rokhlin, S.~Wandzura, The fast multipole method for periodic structures, in:
  Proceedings of the 1994 IEEE Antennas and Propagation Society International
  Symposium, 1994.

\bibitem{li10conf}
S.~Li, V.~Lomakin, Fast interpolation method for field evaluation in a periodic
  unit cell, in: Proceedings of the 2010 IEEE Antennas and Propagation Society
  International Symposium, 2010.

\bibitem{li10}
S.~Li, D.~V. Orden, V.~Lomakin, Fast periodic interpolation method for periodic
  unit cell problems, to appear in IEEE Transactions on Antennas and
  Propagation.

\bibitem{shi10}
Y.~Shi, C.~Chan, Multilevel green's function interpolation method for analysis
  of 3-d frequency selective structures using volume/surface integral equation,
  JOSA A 27 (2010) 308--318.

\bibitem{chen05}
N.~Chen, M.~Lu, F.~Capolino, B.~Shanker, E.~Michielssen, Floquet wave–based
  analysis of transient scattering from doubly periodic, discretely planar,
  perfectly conducting structures, Radio Science 40.

\bibitem{baczewski11}
A.~Baczewski, D.~Dault, B.~Shanker, Accelerated cartesian expansions for the
  rapid solution of periodic electromagnetics problems, submitted to IEEE
  Transactions on Antennas and Propagation.

\bibitem{ewald21}
P.~Ewald, Die berechnung optischer und elektrostatischer gitterpotentiale,
  Annalen der Physik 369 (1921) 253--287.

\bibitem{deleeuw79}
S.~de~Leeuw, J.~Perram, Electrostatic lattice sums for semi-infinite lattices,
  Molecular Physics 37 (1979) 1313--1322.

\bibitem{deleeuw80}
S.~de~Leeuw, J.~Perram, E.~Smith, Simulation of electrostatic systems in
  periodic boundary conditions. i. lattice sums and dielectric constants,
  Proceedings of the Roytal Society of Londong. Series A. 373 (1980) 27--56.

\bibitem{salin00}
G.~Salin, J.~Caillol, Ewald sums for yukawa potentials, Journal of Chemical
  Physics 113 (2000) 10459--10463.

\bibitem{capolino07}
F.~Capolino, D.~Wilton, W.~Johnson, Efficient computation of the 3d green's
  function for the helmholtz operator for a linear array of point sources using
  the ewald method, Journal of Computational Physics 223 (2007) 250--261.

\bibitem{jordan86}
K.~Jordan, G.~Richter, P.~Sheng, An efficient numerical evaluation of the
  green's function for the helmholtz operator on periodic structures, Journal
  of Computational Physics 63 (1986) 222--235.

\bibitem{lovat08}
G.~Lovat, P.~Burghignoli, R.~Araneo, Efficient evaluation of the 3-d periodic
  green's function through the ewald method, 56 56 (2008) 2069--2075.

\bibitem{abramowitz}
M.~Abramowitz, I.~Stegun (Eds.), Handbook of Mathematical Functions, Dover,
  1972.

\bibitem{otani08b}
Y.~Otani, N.~Nishimura, An fmm for periodic boundary value problems for cracks
  for helmholz' equation in 2d, International Journal for Numerical Methods in
  Engineering 73 (2008) 381--406.

\bibitem{chew97}
J.~Son, C.~Lu, W.~Chew, Multilevel fast multipole algorithm for electromagnetic
  scattering by large complex objects, IEEE Transactions on Antennas and
  Propagation 45 (1997) 1488--1493.

\bibitem{lu06}
B.~Lu, X.~Cheng, J.~Huang, J.~McCammon, Order n algorithm for computation of
  electrostatic interactions in biomolecular systems, Proceeds of the National
  Academy of Sciences 103 (2006) 19314--19319.

\bibitem{warren93}
M.~Warren, J.~Salmon, A parallel hased oct-tree n-body algorithm, in:
  Proceedings of the 1993 ACM/IEEE conference on Supercomputing, 1993.

\bibitem{applequist83}
J.~Applequist, Cartesian polytensors, Journal of Mathematical Physics 24 (1983)
  736--742.

\bibitem{walker88}
J.~Walker, Fourier Analysis, Oxford University Press, 1988.

\bibitem{kustepeli00}
A.~Kustepeli, A.~Martin, On the splitting parameter in the ewald method, IEEE
  Transactions on Microwave and Guided Wave Letters 10 (2000) 168--170.

\bibitem{oroskar06}
S.~Oroskar, D.~Jackson, D.~Wilton, Efficient computation of the 2d periodic
  green's function using the ewald method, Journal of Computational Physics 219
  (2006) 899--911.

\end{thebibliography}

\newpage

\begin{table}[ht]
 \centering
 \caption{Convergence of farfield error as a function of the order of ACE expansions, $P$, for 1000 point sources randomly distributed along a line and $\mu=1$.  $\kappa=2\pi$ for Helmholtz and Yukawa potentials, and $|\lv_1|=1$. }
 \label{1d_p_convergence}
 \begin{tabular}{|c|c|c|c|} \hline
  $P$ & $\varepsilon_{far}$ (Helmholtz, $\mu=1$) & $\varepsilon_{far}$ (Yukawa, $\mu=1$) & $\varepsilon_{far}$ (Coulomb, $\mu=1$) \\ \hline
  1 & 2.65262170E-01 & 3.42134128E-01 & 2.01244948E-01 \\ \hline
  3 & 1.86477440E-02 & 7.38115507E-02 & 2.22274736E-02 \\ \hline
  5 & 1.24733083E-03 & 1.30746253E-02 & 3.30542636E-03 \\ \hline
  7 & 6.76186697E-05 & 2.28421189E-03 & 5.58720137E-04 \\ \hline
  9 & 1.42733438E-05 & 4.18448553E-04 & 1.01927872E-04 \\ \hline
 11 & 2.69372066E-06 & 8.04177927E-05 & 1.95815575E-05 \\ \hline
 \end{tabular}
\end{table}

\begin{table}[ht]
 \centering
 \caption{Convergence of farfield error as a function of the order of ACE expansions, $P$, for 1000 point sources randomly distributed over a plane and $\mu=2$.  $\kappa=2\pi$ for Helmholtz and Yukawa potentials, and $|\lv_1|=|\lv_2|=1$. }
 \label{2d_p_convergence}
 \begin{tabular}{|c|c|c|c|} \hline
  $P$ & $\varepsilon_{far}$ (Helmholtz, $\mu=2$) & $\varepsilon_{far}$ (Yukawa, $\mu=2$) & $\varepsilon_{far}$ (Coulomb, $\mu=2$)  \\ \hline
  1 & 2.51998705E-01 & 2.47605092E-01 & 1.43762519E-01 \\ \hline
  3 & 2.26492689E-02 & 3.36631422E-02 & 6.39293888E-03 \\ \hline
  5 & 1.05073461E-03 & 3.70471581E-03 & 6.66751431E-04 \\ \hline
  7 & 3.47767989E-05 & 5.51356326E-04 & 1.21687245E-04 \\ \hline
  9 & 1.43545891E-06 & 8.53940482E-05 & 1.72698036E-05 \\ \hline
 11 & 8.11541488E-07 & 2.82777859E-05 & 7.68416578E-06 \\ \hline
 \end{tabular}
\end{table}

\begin{table}[ht]
 \centering
 \caption{Convergence of farfield error as a function of the order of ACE expansions, $P$, for 1000 point sources randomly distributed over a cube and $\mu=3$.  $\kappa=2\pi$ for Helmholtz and Yukawa potentials, and $|\lv_1|=|\lv_2|=|\lv_3|=1$. }
 \label{3d_p_convergence}
 \begin{tabular}{|c|c|c|c|} \hline
  $P$ & $\varepsilon_{far}$ (Helmholtz, $\mu=3$) & $\varepsilon_{far}$ (Yukawa, $\mu=3$) & $\varepsilon_{far}$ (Coulomb, $\mu=3$) \\ \hline
  1 & 9.23618334E-02 & 2.01036772E-01 & 9.98503574E-02 \\ \hline
  3 & 8.62434445E-03 & 1.58866117E-02 & 1.41604955E-03 \\ \hline
  5 & 4.24578257E-04 & 2.52572255E-03 & 2.66501969E-04 \\ \hline
  7 & 1.71411755E-05 & 8.50629227E-04 & 6.88847400E-05 \\ \hline
  9 & 2.83889405E-06 & 1.88337597E-04 & 2.21046124E-05 \\ \hline
 11 & 2.33785092E-06 & 7.19361796E-05 & 7.88897392E-06 \\ \hline
 \end{tabular}
\end{table}

\begin{table}[ht]
 \centering
 \caption{Convergence of farfield error as a function of the order of ACE expansions, $P$ for 1000 point sources randomly distributed over a cube of unit volume for $\mu=1$ and $\mu=2$.  Both data sets are for a Helmholtz potential with 
$\kappa=2\pi$ and $|\lv_1|=|\lv_2|=1$}
 \label{off_p_convergence}
\begin{tabular}{|c|c|c|} \hline
  $P$ & $\varepsilon_{far}$ (Helmholtz, $\mu=1$) & $\varepsilon_{far}$ (Helmholtz, $\mu=2$) \\ \hline
  1 & 2.24897426E-01 & 2.23438106E-01 \\ \hline
  3 & 2.16007507E-02 & 1.99308663E-02 \\ \hline
  5 & 1.20343605E-03 & 9.69744947E-04 \\ \hline
  7 & 5.26344870E-05 & 2.99860610E-05 \\ \hline
  9 & 1.22053802E-05 & 4.86599084E-06 \\ \hline
 11 & 3.32008243E-06 & 2.67447745E-06 \\ \hline
\end{tabular}
\end{table}

\begin{table}[ht]
 \centering
 \caption{Variation in farfield error as a function of incidence angle for quasi-periodic boundary conditions. For this test, $P=7$ and the geometry consists of 1000 point sources randomly distributed over a plane with $|\lv_1|=|\lv_2|=1$.  
We are concerned with a Helmholtz potential with $\mu=2$ and $\kappa=2\pi$.}
 \label{quasi_error}
 \begin{tabular}{|c|c|c|c|} \hline
  $\theta$ (deg.) & $\varepsilon_{far}$ ($\phi=0$) & $\varepsilon_{far}$ ($\phi=22.5$) & $\varepsilon_{far}$ ($\phi=45$) \\ \hline
  0  & 1.34745750E-05 & 1.34745750E-05 & 1.34745750E-05 \\ \hline
  15 & 1.27088420E-05 & 1.91878307E-05 & 1.76093009E-05 \\ \hline
  30 & 3.61876368E-05 & 5.18579773E-05 & 2.92526545E-05 \\ \hline
  45 & 1.33004484E-05 & 3.65870039E-05 & 1.28715043E-04 \\ \hline
  60 & 1.39020493E-05 & 1.86621752E-05 & 3.71872105E-05 \\ \hline
  75 & 1.60122213E-05 & 2.53068822E-05 & 2.93652039E-05 \\ \hline
  89 & 2.86977094E-05 & 5.52665702E-05 & 7.47714207E-05 \\ \hline
 \end{tabular}
\end{table}

\begin{table}
\centering
\caption{Convergence of farfield error in the $L_2$-norm as the size of the unit cell is increased relative to the wavelength for both periodic and non-periodic ACE.  
Test consists of 1000 point sources randomly distributed over a plane for $\mu=2$ where $|\lv_1|=|\lv_2|=1$, at a fixed number of harmonics (P=8).}
\label{error_degradation}
\begin{tabular}{|c|c|c|} \hline
 $\lambda$ & Free Space & 2D Periodic \\ \hline 
  2.00 & 1.393139705E-005 & 6.799602032E-006 \\ \hline
  1.33 & 1.581618372E-005 & 7.475801890E-006 \\ \hline
  1.00 & 1.704968274E-005 & 1.134334351E-006 \\ \hline
  0.80 & 2.490816540E-005 & 4.231144671E-005 \\ \hline
  0.67 & 1.419596242E-004 & 1.478215770E-004 \\ \hline
  0.57 & 6.760960910E-004 & 2.974738916E-004 \\ \hline
  0.50 & 2.798975397E-003 & 1.003281509E-003 \\ \hline
  0.33 & 2.845979667E-001 & 1.469393475E-001 \\ \hline
  0.25 & 6.951138039E+000 & 1.390723256E+001 \\ \hline
\end{tabular}
\end{table}

%\begin{table}
% \centering
% \caption{Crap} 
% \label{error_degradation}
% \begin{tabular}{|c|c|c|} \hline
% $\lambda$ & Helmholtz & Yukawa  \\ \hline 
%  0.50 	&  2.191368373E-002  &  1.547398530E-003  \\ \hline
%  0.75	&  4.639230288E-004  &  4.782700849E-004  \\ \hline
%  1	&  8.762108743E-005  &  2.379302209E-004  \\ \hline
%  2	&  3.511440571E-005  &  6.556593996E-005  \\ \hline
%  4	&  1.902226099E-005  &  2.488033862E-005  \\ \hline
%  8	&  9.681553088E-006  &  1.095262669E-005  \\ \hline
%  16	&  4.861844154E-006  &  5.156782454E-006  \\ \hline
%  32	&  2.433471593E-006  &  2.504434082E-006  \\ \hline
%  64	&  1.217003017E-006  &  1.234402160E-006  \\ \hline
%  128	&  6.086381471E-007  &  6.129562108E-007  \\ \hline
%  256	&  3.042983255E-007  &  3.053649428E-007  \\ \hline
%  512	&  1.521769250E-007  &  1.524447526E-007  \\ \hline
%  1024	&  7.639826086E-008  &  7.642419020E-008  \\ \hline
% \end{tabular}
%\end{table}

\begin{table}[ht]
 \centering
 \caption{Scaling with the number of unknowns, $N$ for Yukawa point sources randomly distributed over a plane with $\mu=1$.  A linear regression on a log-log scale indicates $t_{ACE} \sim N^{1.029}$ and $N_{near} \sim N^{1.030}$.}
 \label{1d_scaling}
 \begin{tabular}{|c|c|c|c|c|c|c|} \hline
   $N$ ($N_l$) & $t_{ACE}$ (sec) & $t_{pre}$ (sec) & $N_{unique}$ & $t_{far}$ (sec) & $t_{direct}$ (sec) & $N_{near}$ \\ \hline
   1024 (3) & 2.00E-02 & 1.16E-00 & 34  & 1.46E+01  	  & 2.00E-02 	   & 495728  \\ \hline
   4096 (4) & 1.60E-01 & 3.75E-00 & 110 & 3.70E+02 	  & 6.50E-01 	   & 2188068 \\ \hline
  16384 (5) & 5.20E-01 & 5.98E-00 & 176 & $\sim$6.57E+03  & $\sim$1.15E+01 & 9277942 \\ \hline
  65536 (6) & 2.01E-00 & 8.22E-00 & 242 & $\sim$1.08E+05  & $\sim$1.90E+02 & 38303868 \\ \hline
 262144 (7) & 7.97E-00 & 1.00E+01 & 297 & $\sim$1.74E+06  & $\sim$3.05E+03 & 156415674 \\ \hline
1048576 (8) & 3.19E+01 & 1.12E+01 & 333 & $\sim$2.79E+07  & $\sim$4.90E+04 & 634053672 \\ \hline   
 \end{tabular}
\end{table}

%\begin{table}[ht]
% \centering
% \caption{Scaling with the number of unknowns, $N$ for Yukawa point sources randomly distributed over a plane with $\mu=1$.  A linear regression on a log-log scale indicates $t_{ACE} \sim N^{1.029}$ and $N_{near} \sim N^{1.030}$.}
% \label{1d_scaling}
% \begin{tabular}{|c|c|c|c|c|c|c|} \hline
%   $N$ ($N_l$) & $t_{ACE}$ (sec) & $t_{pre}$ (sec) & $N_{unique}$ & $t_{far}$ (sec) & $t_{direct}$ (sec) & $N_{near}$ \\ \hline
%   1024 (3) & 1.9999E-02 & 1.1600E-00 & 34  & 1.4640E+01  	  & 1.9999E-02 	     & 495728  \\ \hline
%   4096 (4) & 1.5999E-01 & 3.7500E-00 & 110 & 3.6950E+02 	  & 6.4999E-01 	     & 2188068 \\ \hline
%  16384 (5) & 5.2000E-01 & 5.9800E-00 & 176 & $\sim$6.5716E+03    & $\sim$1.1546E+01 & 9277942 \\ \hline
%  65536 (6) & 2.0099E-00 & 8.2199E-00 & 242 & $\sim$1.0794E+05	  & $\sim$1.8964E+02 & 38303868 \\ \hline
% 262144 (7) & 7.9699E-00 & 1.0030E+01 & 297 & $\sim$1.7386E+06    & $\sim$3.0547E+03 & 156415674 \\ \hline
%1048576 (8) & 3.1919E+01 & 1.1239E+01 & 333 & $\sim$2.7865E+07    & $\sim$4.8958E+04 & 634053672 \\ \hline   
% \end{tabular}
%\end{table}

\begin{table}[ht]
 \centering
 \caption{Scaling with the number of unknowns, $N$, for Helmholtz point sources randomly distributed over a plane with $\mu=2$.  A linear regression on a log-log scale indicates $t_{ACE} \sim N^{1.028}$ and $N_{near} \sim N^{1.007}$.}
 \label{2d_scaling}
 \begin{tabular}{|c|c|c|c|c|c|c|} \hline
   $N$ ($N_l$) & $t_{ACE}$ (sec) & $t_{pre}$ (sec) & $N_{unique}$ & $t_{far}$ (sec) & $t_{direct}$ (sec) & $N_{near}$ \\ \hline
   1024 (3) & 2.00E-02 & 1.50E-01 & 24  & 8.63E+01  	  & 2.00E-02       & 590526  \\ \hline
   4096 (4) & 1.70E-01 & 8.90E-01 & 168 & 2.62E+03 	  & 6.40E-01       & 2364674 \\ \hline
  16384 (5) & 5.60E-01 & 1.50E-00 & 288 & $\sim$4.97E+04  & $\sim$1.15E+01 & 9511212 \\ \hline
  65536 (6) & 2.14E-00 & 2.08E-00 & 408 & $\sim$7.72E+05  & $\sim$1.89E+02 & 38504020 \\ \hline
 262144 (7) & 8.24E-00 & 2.44E-00 & 528 & $\sim$1.24E+07  & $\sim$3.04E+03 & 156415674 \\ \hline
1048576 (8) & 3.23E+01 & 2.60E-00 & 592 & $\sim$1.99E+08  & $\sim$4.88E+04 & 634053672 \\ \hline   
 \end{tabular}
\end{table}

%\begin{table}[ht]
% \centering
% \caption{Scaling with the number of unknowns, $N$, for Helmholtz point sources randomly distributed over a plane with $\mu=2$.  A linear regression on a log-log scale indicates $t_{ACE} \sim N^{1.028}$ and $N_{near} \sim N^{1.007}$.}
% \label{2d_scaling}
% \begin{tabular}{|c|c|c|c|c|c|c|} \hline
%   $N$ ($N_l$) & $t_{ACE}$ (sec) & $t_{pre}$ (sec) & $N_{unique}$ & $t_{far}$ (sec) & $t_{direct}$ (sec) & $N_{near}$ \\ \hline
%   1024 (3) & 1.9999E-02 & 1.5000E-01 & 24  & 8.6329E+01  	  & 1.9999E-02 	     & 590526  \\ \hline
%   4096 (4) & 1.6999E-01 & 8.8999E-01 & 168 & 2.6151E+03 	  & 6.3999E-01 	     & 2364674 \\ \hline
%  16384 (5) & 5.5999E-01 & 1.5000E-00 & 288 & $\sim$4.9681E+04    & $\sim$1.1497E+01 & 9511212 \\ \hline
%  65536 (6) & 2.1399E-00 & 2.0800E-00 & 408 & $\sim$7.7233E+05	  & $\sim$1.8900E+02 & 38504020 \\ \hline
% 262144 (7) & 8.2400E-00 & 2.4399E-00 & 528 & $\sim$1.2440E+07	  & $\sim$3.0445E+03 & 156415674 \\ \hline
%1048576 (8) & 3.2309E+01 & 2.5999E-00 & 592 & $\sim$1.9939E+08 	  & $\sim$4.8795E+04 & 634053672 \\ \hline   
% \end{tabular}
%\end{table}

\begin{table}[ht]
 \centering
 \caption{Scaling with the number of unknowns, $N$, for Coulomb point sources randomly distributed over a cube with $\mu=3$.  A linear regression on a log-log scale indicates $t_{ACE} \sim N^{1.030}$ and $N_{near} \sim N^{1.002}$.}
 \label{3d_scaling}
 \begin{tabular}{|c|c|c|c|c|c|c|} \hline
   $N$ ($N_l$) & $t_{ACE}$ (sec) & $t_{pre}$ (sec) & $N_{unique}$ & $t_{far}$ (sec) & $t_{direct}$ (sec) & $N_{near}$ \\ \hline
   4096  (3) & 2.10E-01 & 7.26E-00 & 218  & 2.25E+03  	      & 6.40E-01       & 7081842  \\ \hline
   32768 (4) & 6.76E-00 & 3.91E+01 & 2290 & $\sim$2.36E+05    & $\sim$6.71E+02 & 56904834 \\ \hline
  262144 (5) & 2.49E+01 & 5.32E+01 & 3678 & $\sim$1.58E+07    & $\sim$4.51E+03 & 458747022 \\ \hline
 2097152 (6) & 1.72E+02 & 6.62E+01 & 5066 & $\sim$1.02E+09    & $\sim$2.90E+05 & 3724629570 \\ \hline
 \end{tabular}
\end{table}

%\begin{table}[ht]
% \centering
% \caption{Scaling with the number of unknowns, $N$, for Coulomb point sources randomly distributed over a cube with $\mu=3$.  A linear regression on a log-log scale indicates $t_{ACE} \sim N^{1.030}$ and $N_{near} \sim N^{1.002}$.}
% \label{3d_scaling}
% \begin{tabular}{|c|c|c|c|c|c|c|} \hline
%   $N$ ($N_l$) & $t_{ACE}$ (sec) & $t_{pre}$ (sec) & $N_{unique}$ & $t_{far}$ (sec) & $t_{direct}$ (sec) & $N_{near}$ \\ \hline
%   4096  (3) & 2.0999E-01 & 7.2599E-00 & 218  & 2.2513E+03  	    & 6.4000E-01       & 7081842  \\ \hline
%   32768 (4) & 6.7599E-00 & 3.9139E+01 & 2290 & $\sim$2.3611E+05    & $\sim$6.7121E+02 & 56904834 \\ \hline
%  262144 (5) & 2.4929E+01 & 5.3199E+01 & 3678 & $\sim$1.5850E+07    & $\sim$4.5058E+03 & 458747022 \\ \hline
% 2097152 (6) & 1.7199E+02 & 6.6169E+01 & 5066 & $\sim$1.0203E+09    & $\sim$2.9006E+05 & 3724629570 \\ \hline
% \end{tabular}
%\end{table}

\end{document}